    \documentclass[12pt,journal,onecolumn,draftcls]{IEEEtran}

\usepackage[utf8]{inputenc} 
\usepackage[T1]{fontenc}
\usepackage{url}
\usepackage{ifthen}
\usepackage{cite}
\usepackage[cmex10]{amsmath} 
                             \usepackage{epsf}
\usepackage{xcolor}

\usepackage{amssymb}
\usepackage{epsfig,verbatim}
\usepackage{algorithm}
\usepackage{graphicx}
\usepackage{tikz}
\usepackage{pgfplots}
\usepackage{tkz-euclide}
\usepackage{tikz-3dplot}

\usepackage[center]{caption}

\usepackage{subcaption}
\usepackage{cleveref}

\usepackage{algpseudocode}
\long\def\comment#1{}

\newfont{\bbb}{msbm10 scaled 700}

\newfont{\bb}{msbm10 scaled 1100}

\newcommand{\bv}{{\bf b}}
\newcommand{\cv}{{\bf c}}

\newcommand{\ev}{{\bf e}}

\newcommand{\nv}{{\bf n}}

\newcommand{\qv}{{\bf q}}

\newcommand{\tv}{{\bf t}}

\newcommand{\vv}{{\bf v}}
\newcommand{\xv}{{\bf x}}
\newcommand{\yv}{{\bf y}}
\newcommand{\zv}{{\bf z}}

\newcommand{\Gm}{{\bf G}}

\newcommand{\Nm}{{\bf N}}

\newcommand{\Xm}{{\bf X}}
\newcommand{\Ym}{{\bf Y}}
\newcommand{\Zm}{{\bf Z}}

\newcommand{\Ac}{{\cal A}}
\newcommand{\Bc}{{\cal B}}

\newcommand{\Ec}{{\cal E}}

\newcommand{\Hc}{{\cal H}}

\newcommand{\Nc}{{\cal N}}
\newcommand{\Oc}{{\cal O}}
\newcommand{\Pc}{{\cal P}}
\newcommand{\Qc}{{\cal Q}}
\newcommand{\Rc}{{\cal R}}
\newcommand{\Sc}{{\cal S}}

\newcommand{\Wc}{{\cal W}}

\newcommand{\nuv}{\hbox{\boldmath$\nu$}}
\newcommand{\muv}{\hbox{\boldmath$\mu$}}

\renewcommand{\arg}{{\hbox{arg}}}

\usepackage{times}
\usepackage{amsthm}

\newtheorem{theorem}{Theorem}
\newcommand{\thmlabel}[1]{\label{thm:#1}}
\newcommand{\thmref}[1]{\ref{thm:#1}}

\newtheorem{definition}{Definition}

\newtheorem{lemma}{Lemma}

\newtheorem{remark}{Remark}

\newtheorem{proposition}{Proposition}

\newcommand{\argmax}{\operatornamewithlimits{argmax}}
\newcommand{\argmin}{\operatornamewithlimits{argmin}}

\newsavebox{\smlmat}
\savebox{\smlmat}{$\left[\begin{smallmatrix}1& 0 & 0 \\ 0 &1 & 0 \\ 0 & 0 & 2\end{smallmatrix}\right]$}

\begin{document}
\title{Recovering Data Permutations from Noisy Observations: The Linear Regime} 

\IEEEoverridecommandlockouts  


 \author{%
   \IEEEauthorblockN{Minoh Jeong\IEEEauthorrefmark{1},
                     Alex Dytso\IEEEauthorrefmark{2},
                     Martina Cardone\IEEEauthorrefmark{1},
                     and H. Vincent Poor\IEEEauthorrefmark{2}\\}
   \IEEEauthorblockA{\IEEEauthorrefmark{1}%
                     University of Minnesota,
                    Minneapolis, MN 55455, USA,
                     \\\{jeong316, mcardone\}@umn.edu}
 \\  \IEEEauthorblockA{\IEEEauthorrefmark{2}%
                     Princeton University,
                    Princeton, NJ 08544, USA,
                    \\ \{adytso, poor\}@princeton.edu}
                     \thanks{The work  of M.~Jeong and M.~Cardone was supported in part by the U.S. National Science Foundation under Grant CCF-1849757. {The work  of A.~Dytso and H. V. Poor was supported in part by the U.S. National Science Foundation under Grant CCF-1908308. Part of this work was presented at the 2020 IEEE International Symposium on Information Theory~\cite{jeong20}.}}

 }

\maketitle

\begin{abstract}
This paper considers a noisy data structure recovery problem. The goal is to investigate the following question: Given a noisy observation of a permuted data set, according  to which permutation was the original data sorted?
The focus is on scenarios where data is generated according to an isotropic Gaussian distribution, and the noise is additive Gaussian with an arbitrary  covariance matrix.
This problem is posed within a hypothesis testing framework. 
The objective is to study the {\em linear} regime in which the optimal decoder has a polynomial complexity in the data size, and it declares the permutation by simply computing a permutation-independent linear function of the noisy observations.

The main result of the paper is a complete characterization of the linear regime in terms of the noise covariance matrix. Specifically, it is shown that this matrix must have a very flat spectrum with at most three distinct eigenvalues to induce the linear regime. Several practically relevant implications of this result are discussed, and the error probability incurred by the decision criterion in the linear regime is also characterized. A core technical component consists of using linear algebraic and geometric tools, such as  Steiner symmetrization.
\end{abstract}



\section{Introduction}
The problem of recovery of the original permutation from noisy permuted data is a common task in modern communication and computing systems. For example, in the data analytics realm, recommender systems are often more interested in recovering the relative ranking of data points rather than the values of the data itself. Furthermore, users may desire to privatize their data before it is collected from an external party. A suitable solution to privatize data and hence maintain its confidentiality consists of perturbing it with noise. Upon receiving the perturbed/noisy data, the recommender system will then need to recover the data permutation (e.g., ranking of users' interests) in order to provide the next recommendation.

In this work, we investigate the following question on noisy data structure recovery: Given a noisy observation of a permuted data set, according  to which permutation was the original data sorted?

\subsection{Related Work}
Data  permutation recovery has recently gained significant importance, and it is a problem studied in various fields~\cite{Collier,Dytso2019,Pananjady2018,Pananjady2017,Slawski2019,abid2017linear,abid2018stochastic,Emiya2014,Shi2020,Wang2018,slawski2019sparse}.
For instance, in the machine learning literature, the problem of feature matching in computer vision is often formulated as a permutation estimation problem~\cite{Collier}.
In particular, the goal of~\cite{Collier} is to estimate the permutation that matches two sets of features given noisy observations. 
As another example, in~\cite{Dytso2019} the authors propose a framework to estimate the values of an original sorted vector, given a noisy sorted observation of it. They show that, under certain symmetry conditions, the minimum mean square error estimator can be characterized by a linear combination of estimators on the unsorted data. 


Studies on the permutation recovery problem have also recently appeared in linear regression. In~\cite{Pananjady2018}, the authors analyze the permutation recovery problem and consider an output given by an input that is permuted by an unknown permutation matrix.
They provide necessary and sufficient conditions on the signal-to-noise ratio for exact permutation recovery. The multivariate linear regression model with unknown permutation is studied in~\cite{Pananjady2017}. The authors characterize the minimax prediction error and analyze estimators. 
A similar model with sparsely permuted data can be found in~\cite{Slawski2019}. 
A study on isotonic regression without data labels, namely the {\em uncoupled isotonic regression}, is discussed in~\cite{rigollet2019uncoupled}. 
In particular, the goal consists of estimating a non-decreasing regression function given unordered sets of data.
A study on the seriation problem, where the goal is to estimate a pair of unknown permutation and data matrices from a noisy observation, can be found in~\cite{flammarion2019optimal}.

Estimating data given randomly selected measurements -- which is termed {\em unlabeled sensing} -- is studied in \cite{Unnikrishnan2018,Haghighatshoar2018_2,Zhang2019}. 
A necessary condition on the dimension of the observation vector for uniquely recovering the original data in the noiseless case is provided in~\cite{Unnikrishnan2018}.
Design and discussion on recovery algorithms can be found in~\cite{Haghighatshoar2018_2,Hsu2017,saab2019shuffled}. A generalization of the framework in~\cite{Unnikrishnan2018} is provided in~\cite{dokmanic2019permutations} and it considers any invertible and diagonalizable matrix rather than the classical permutation (selection) matrix.
The authors in~\cite{tsakiris19a} and~\cite{tsakiris2018eigenspace} propose a framework - referred to as {\em homomorphic sensing} - that encompasses the unlabeled sensing framework in~\cite{Unnikrishnan2018}.

Applications of permutation recovery on the biostatistics area can be found in~\cite{Ma2019}. In particular, the exact and partial recoveries for the microbiome growth dynamics are discussed. Further, in~\cite{Marano2019} the authors characterize the fundamental limit for the performance of a hypothesis testing problem with unknown labels, and they propose suitable algorithms for the problem.

\subsection{Contributions}
In this paper, we investigate the noisy data permutation recovery problem, which consists of recovering the permutation of an original data vector of size $n$ that has been perturbed by noise.
We consider a scenario where data is generated according to an isotropic Gaussian distribution, and the perturbation consists of adding Gaussian noise that can have an arbitrary covariance matrix, i.e., noise can have memory. 
Our main contributions can be summarized as follows:
\begin{enumerate}
\item We formulate the problem within a hypothesis testing framework, which consists of $n!$ hypotheses.
The optimal decision criterion for the hypothesis testing problem is given by the celebrated Neyman-Pearson lemma, which formulates the optimal decision regions in terms of a ratio of some likelihood functions. We show that the optimal decision regions of the considered hypothesis testing problem must have a certain symmetry.

\item We show that the optimal decision regions may or may not be a linear transformation of the corresponding hypothesis regions depending on the noise covariance matrix.
We focus our study on the {\em linear regime} where the optimal permutation decoding consists of a simple  linear transformation of the noisy observation, followed by a sorting algorithm outputting the permutation along which this linear transformation is sorted. The computed linear transformation is the same across all  permutations and hence, throughout the paper we refer to it as permutation-independent. 
This regime is particularly appealing as within it the optimal decoder has a complexity that is at most polynomial in $n$, as opposed to a brute force approach that would incur a computational complexity of $n!$.

\item We characterize the optimal decision criterion for the hypothesis testing problem in the linear regime, by deriving the optimal decision regions. 
 In particular, we show that the optimal decoder declares the permutation based only on a permutation-independent linear function of the noisy observation. 
Our result provides both a {\em linear algebraic} and a {\em geometric} interpretations of the linear regime in terms of the noise covariance matrix. Specifically, the linear algebraic viewpoint says that the noise covariance matrix can have at most three distinct eigenvalues. The geometric interpretation, instead says that the $n$-dimensional ellipsoid, characterized by a function of the noise covariance matrix, when projected onto a specific hyperplane has to be an $(n-1)$-dimensional ball.
To derive these results, a core technical component consists of using linear algebraic and geometric tools, such as the Schur complement and Steiner symmetrization.

\item With the structure of the optimal decision regions in the linear regime, we discuss several practically relevant implications and
special cases. 
 For instance, we prove that when $n=2$ the linear regime is the only regime. 
For the class of diagonal noise covariance matrices and $n>2$, we show that the noise covariance matrix must have all equal diagonal elements to fall within the linear regime, i.e, if the noise is memoryless, then it must be isotropic.
Finally, we characterize the probability of error incurred by the decision criterion in the linear regime.
In particular, we express the probability of error in terms of the volume of a region which consists of the intersection of a cone with a permutation-independent linear transformation of the unit radius $2n$-dimensional ball.
\end{enumerate}

\subsection{Paper Organization}
Section~\ref{sec:PrForm} introduces the notation and formulates the hypothesis testing problem. 
Section~\ref{sec:Optimal_Region} discusses the optimal decision regions for our hypothesis testing problem. 
Section~\ref{sec:MainRes} provides the main result of the paper, which consists of the characterization of the optimal decision regions in the linear regime. Section~\ref{sec:MainRes} also discusses several implications of the main result.
Section~\ref{sec:ProofMainResu} provides a detailed proof of the main result.
Finally, Section~\ref{sec:Conclusion} concludes the paper. 
Some of the proofs can be found in the appendix.  The paper contains several 3D figures, the interactive versions of which can be found in \cite{gitHub}.

\section{Notation and Problem Formulation}
\label{sec:PrForm}
\noindent{\bf{Notation.}} Boldface upper case letters $\mathbf{X}$ denote vector random variables; the boldface lower case letter $\mathbf{x}$ indicates a specific realization of $\mathbf{X}$; 
$[n_1: n_2]$ is the set of integers from $n_1$ to $n_2 \geq n_1$;
$I_n$ is the identity matrix of dimension $n$; 
$\mathbf{0}_n$ (respectively, $\mathbf{1}_n$) is the column vector of dimension $n$ of all zeros (respectively, ones); 
$0_{n \times k}$ (respectively, $1_{n \times k}$) is an $n \times k$ matrix of all zeros (respectively, ones); 
$\text{det}(A)$ is the determinant of the matrix $A$;
$\|\mathbf{x}\|$ is the $\ell_2$ norm of $\mathbf{x}$, and $\mathbf{x}^T$ is the transpose of $\mathbf{x}$.
Calligraphic letters indicate sets;
$|\mathcal{A}|$ denotes the cardinality of the set $\mathcal{A}$; for two sets $\mathcal{A}$ and $\mathcal{B}$, 
$\mathcal{A} \cap \mathcal{B}$ is the set of elements that belong both to $\mathcal{A}$ and $\mathcal{B}$;
$\varnothing$ is the empty set.
For a set $\mathcal{S} \subseteq \mathbb{R}^k$, $\mathrm{Vol}^k(\mathcal{S})$ denotes the volume, i.e., the $k$-dimensional Lebesgue measure, of $\mathcal{S}$;
$\mathcal{B}^n(\cv,r)$ denotes the $n$-dimensional ball centered at $\cv \in \mathbb{R}^n$ with radius $r$.
Finally, the multiplication of a matrix $A$ by a set $\mathcal{B}$ is denoted and defined as $A \mathcal{B}= \{ Ax: x \in   \mathcal{B} \}  $.  \hfill$\square$

   \begin{figure*}
        \centering
        \includegraphics[width=\textwidth]{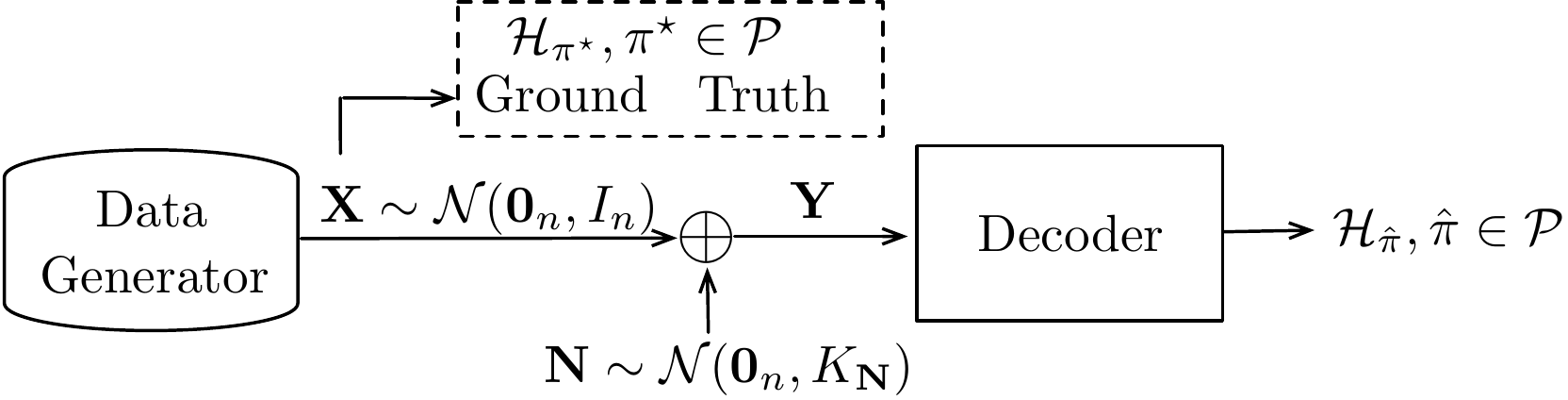}
        \caption{Graphical representation of the proposed framework.}
        \label{fig:Framework}
    \end{figure*}

\medskip

We consider the framework in Fig.~\ref{fig:Framework}, where an $n$-dimensional random vector $\mathbf{X}$ is generated according to an isotropic Gaussian distribution, namely $\mathbf{X} \sim \mathcal{N}(\mathbf{0}_n, {I}_n)$. The random vector $\mathbf{X}$ is then passed through an additive Gaussian noise channel, the output of which is denoted as $\mathbf{Y}$. In other words, we have
$\mathbf{Y} = \mathbf{X} + \mathbf{N}$, with $\mathbf{N} \sim \mathcal{N}(\mathbf{0}_n,K_{\mathbf{N}})$ where $K_{\mathbf{N}}$ denotes the covariance matrix of the additive noise $\mathbf{N}$, and where $ \mathbf{X}$ and $\mathbf{N}$ are independent. 

In this work, we are interested in answering the following question: 
Given the observation of $\mathbf{Y}$, according to which permutation - among the $n!$ possible ones - was the vector $\mathbf{X}$ sorted?
Towards this end, we define $\mathcal{P}$ as the collection of all permutations of the elements of $[1:n]$; clearly $|\mathcal{P}| = n!$. We formulate a hypothesis testing
problem with $n!$ hypotheses $\mathcal{H}_{\pi}, \pi \in \mathcal{P}$, where $\mathcal{H}_{\pi}$ is the hypothesis that $\mathbf{X}$ is an $n$-dimensional vector sorted according to the permutation $\pi \in \mathcal{P}$.
Formally, each hypothesis corresponds to the following set,
\begin{align}
\Hc_\pi = \{\mathbf{x} \in \mathbb{R}^n:x_{\pi_{1}} \le x_{\pi_2} \le \cdots \le x_{\pi_n} \},
\end{align} 
where $x_{\pi_i}, i \in [1:n]$ is the $\pi_i$-th element of $\mathbf{x}$, and $\pi_i, i \in[1:n]$ is the $i$-th element of $\pi$.
Note that the hypotheses $\mathcal{H}_{\pi}$'s divide the entire $n$-dimensional space into $n!$ regions -- referred to as {\em hypothesis regions} -- and each hypothesis is associated to one of these regions.  Moreover, due to the symmetry of $\mathbf{X}$ we have that  $\Pr \left ( \mathbf{X}  \in \mathcal{H}_{\pi} \right ) =\frac{1}{n!}, \forall \pi \in  \mathcal{P}$.  

We seek to characterize the {\em optimal} decision criterion among the $n!$ hypotheses.
In other words, with reference to Fig.~\ref{fig:Framework}, we are interested in characterizing the  {\em decision rule (decoder)}, so that its output $\mathcal{H}_{\hat{\pi}}, \hat{\pi} \in \mathcal{P}$ is such that
\begin{align}
\label{eq:OptCrit}
\mathcal{H}_{\hat{\pi}}: \hat{\pi} = \argmin_{\pi \in \mathcal{P}} \ \{ \Pr \left (\mathcal{H}_{\pi} \neq \mathcal{H}_{\pi^\star}\right ) \},
\end{align}
where $\pi^\star$ denotes the permutation according to which the random vector $\mathbf{X}$ is sorted.

\begin{figure}
	\centering
	\includegraphics[width=280pt]{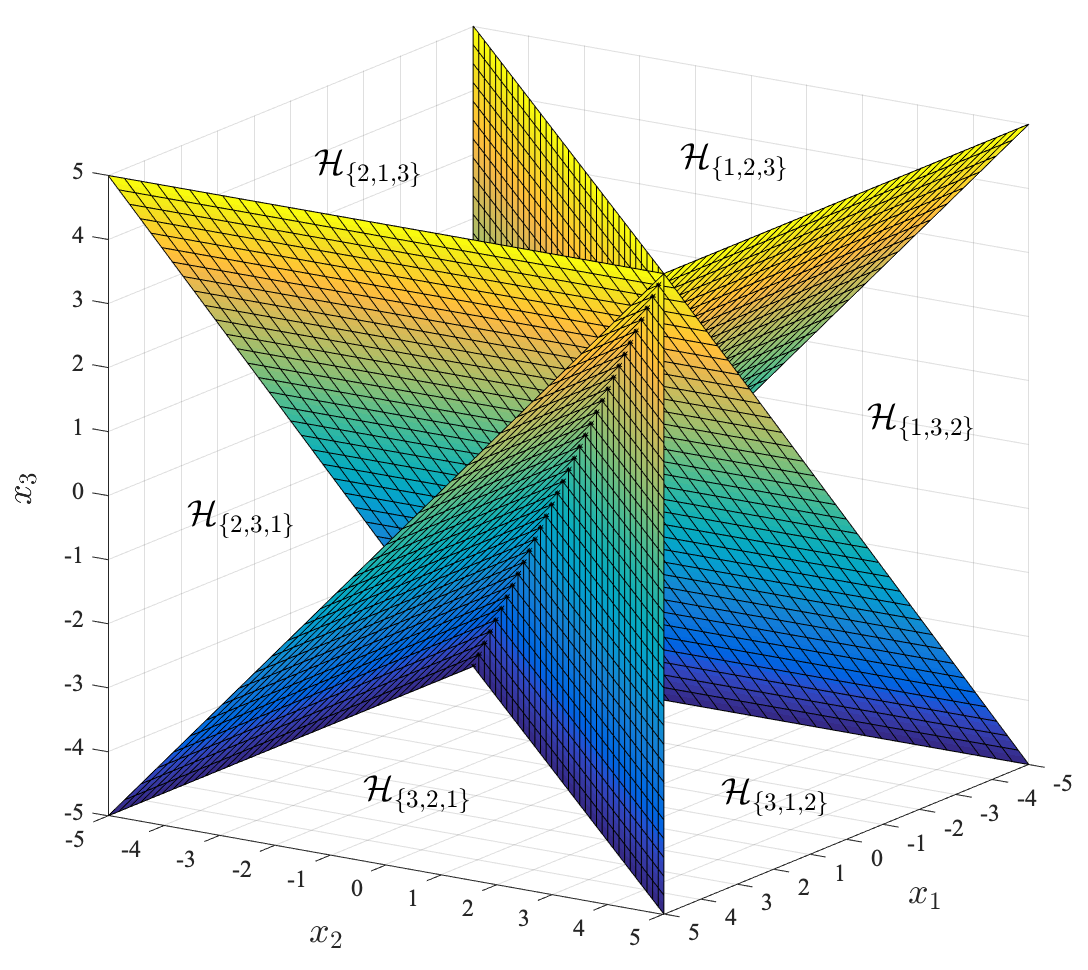}
	\caption{Case $n=3$. Graphical representation of the hypothesis regions associated to each of the $6$ hypotheses.}
	\label{fig:RegN3}
\end{figure}

\noindent {\bf{Example.}} Let $n=3$, then we have $|\mathcal{P}| = 6$ and hypotheses $\mathcal{H}_{\pi}, \pi \in \mathcal{P}$ defined as
\begin{align*}
& \mathcal{H}_{\{1,2,3\}}: X_1 \leq X_2 \leq X_3, \quad  \mathcal{H}_{\{1,3,2\}}: X_1 \leq X_3 \leq X_2,
\\ & \mathcal{H}_{\{2,1,3\}}: X_2 \leq X_1 \leq X_3, \quad  \mathcal{H}_{\{2,3,1\}}: X_2 \leq X_3 \leq X_1,
\\ & \mathcal{H}_{\{3,1,2\}}: X_3 \leq X_1 \leq X_2, \quad  \mathcal{H}_{\{3,2,1\}}: X_3 \leq X_2 \leq X_1,
\end{align*}
where $X_i, i \in [1:3]$ is the $i$-th element of $\mathbf{X}$.
Each hypothesis is hence associated to a hypothesis region in the $3$-dimensional space, as also graphically represented in Fig.~\ref{fig:RegN3}.

\section{Optimal Decision Regions}
\label{sec:Optimal_Region}
In this section, using standard hypothesis testing tools we characterize the optimal decision criterion. We also make general statements about the structure of the decision regions. 
Towards this end, we make use of the result in~\cite[Appendix~3C]{Kay1998}, which shows that, for an observation $\yv$, the  
%
optimal decision criterion in~\eqref{eq:OptCrit}  is given by  the maximum a posterior probability (MAP) decoder, namely
\begin{subequations}
\label{test1}
\begin{align}
&
\mathcal{H}_{\hat{\pi}}: \ \hat{\pi}=\argmax_{\pi \in \mathcal{P}} \ \{f_\Ym(\yv,\mathcal{H}_{\pi})\},
\\ & f_\Ym(\yv,\mathcal{H}_{\pi}) = f_\Ym(\yv|\mathcal{H}_{\pi}) \Pr(\mathcal{H}_{\pi}), \ \pi \in \mathcal{P},\label{eq:hypodensity}
\end{align}
\end{subequations}
where $f_\Ym(\yv|\mathcal{H}_{\pi})$ denotes the conditional probability density function (PDF) of $\Ym$ given that $\Xm \in \mathcal{H}_{\pi}$.
By defining the likelihood functions $L(\yv,\mathcal{H}_{\pi})=f_\Ym(\yv|\mathcal{H}_{\pi}), \forall \pi \in \mathcal{P}$, we have that~\eqref{test1} can be equivalently formulated as
\begin{equation}\label{test}
\mathcal{H}_{\hat{\pi}}: \
 {L(\yv,\mathcal{H}_{\hat{\pi}})\over L(\yv,\mathcal{H}_{\pi})} \geq 1	,~\forall \pi \ne \hat{\pi},
\end{equation} 
where we have used the fact that $\Pr(\mathcal{H}_{\pi}) = \Pr(\mathcal{H}_{\tau}), \forall (\pi,\tau) \in \mathcal{P} \times \mathcal{P}$, which follows since $\mathbf{X} \sim \mathcal{N}(\mathbf{0}_n, {I}_n)$.
It is worth noting that, since $\Xm$ and $\mathbf{N}$ are independent, then the likelihood function $L(\yv,\mathcal{H}_{\pi}), \pi \in \mathcal{P}$  can be expressed by using the convolution between two PDFs as
\begin{equation}
\label{eq:likelCon}
L(\yv,\mathcal{H}_{\pi}) = \mathbb{E}\left[f_\mathbf{N}(\yv-\Xm)|\mathcal{H}_{\pi} \right],
\end{equation} 
where $f_\mathbf{N}(\cdot)$ is the PDF of $\mathbf{N}$.

\begin{figure}
	\centering
	\includegraphics[width=0.8\textwidth]{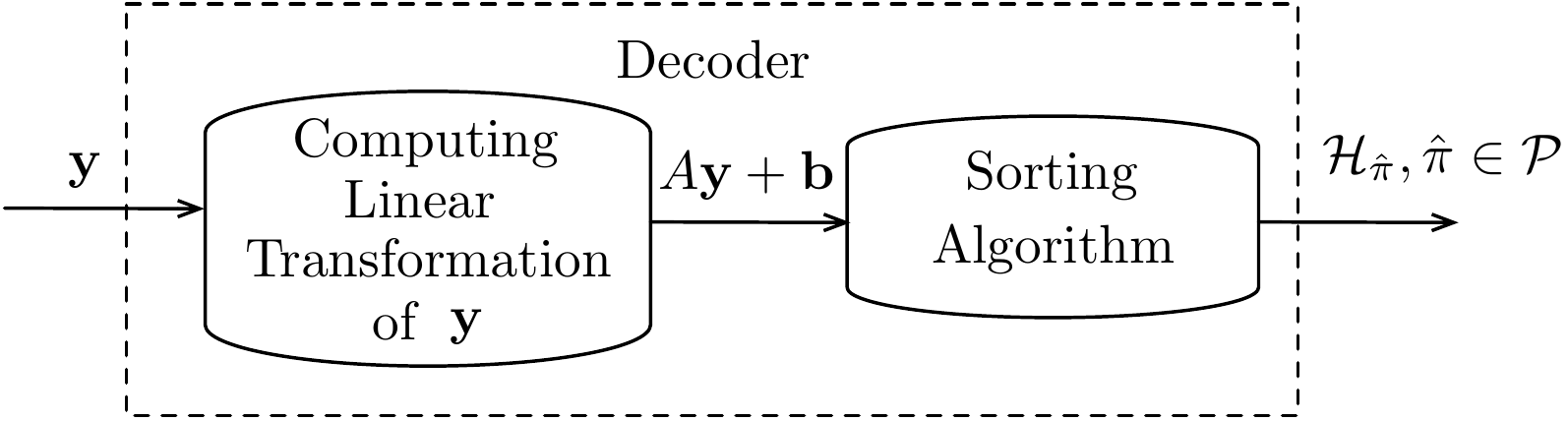}
	\caption{Block diagram of the optimal  decoder in the linear regime. }
	\label{fig:Block}
\end{figure}

With the formulation in~\eqref{test}, we can now define the {\em optimal} decision regions $\mathcal{R}_{\pi,K_{\mathbf{N}}}, \pi \in \mathcal{P}$ of our hypothesis testing problem\footnote{The notation $\mathcal{R}_{\pi,K_{\mathbf{N}}}$ indicates that, in general, the decision regions  might be functions of the noise covariance matrix $K_{\mathbf{N}}$.}.
In particular, the decision criterion will leverage these regions to output $\mathcal{H}_{\hat{\pi}}, \hat{\pi} \in \mathcal{P}$, namely if the observation vector $\mathbf{y} \in \mathcal{R}_{\pi,K_{\mathbf{N}}}$, then the  decoder would declare that the input vector $\mathbf{x} \in \mathcal{H}_{\pi}$.
We have that the optimal decision region $\mathcal{R}_{\pi,K_{\mathbf{N}}}$ corresponding to the hypothesis region $\mathcal{H}_{\pi}, \pi \in \mathcal{P}$ is defined as
\begin{align}\label{region}
\mathcal{R}_{\pi,K_{\mathbf{N}}}
&=\left\{\yv \in \mathbb{R}^n:f_\Ym(\yv,\mathcal{H}_{\pi})  \geq  \max_{\substack{\tau \in \mathcal{P} \\ \tau \neq \pi}}f_\Ym(\yv,\mathcal{H}_{\tau}) \right\} \nonumber \\ 
& = \left\{\yv \in \mathbb{R}^n:{L(\yv,\mathcal{H}_{\pi})\over L(\yv,\mathcal{H}_{\tau})} \geq 1,~\forall \tau \in \mathcal{P}, \tau \ne \pi \right\}.
\end{align} 
\begin{remark}
If $\mathbf{y} \in \mathbb{R}^n$ belongs to the boundary between two or more decision regions,  then we arbitrarily select one of the $\mathcal{H}_{\pi}, \pi \in \mathcal{P}$ associated to these candidate decision regions.
\end{remark}
The objective of this work is to characterize \emph{sufficient and necessary} conditions on the noise covariance matrix  $K_{\mathbf{N}}$ such that each optimal decision region $\mathcal{R}_{\pi,K_{\mathbf{N}}}, \pi \in \mathcal{P}$ in~\eqref{region} is a permutation-independent {\em linear} transformation of the corresponding hypothesis region $\mathcal{H}_{\pi}$ (i.e., $\mathcal{R}_{\pi,K_{\mathbf{N}}}=A \Hc_{\pi}+{\bf b}$ for some $A \in \mathbb{R}^{n \times n}$ and ${\bf b} \in \mathbb{R}^n$, which are the same across all permutations).  In other words, we seek to characterize the regime in which the optimal 
decoder
consists of computing a simple permutation-independent linear transformation of the noisy observation $\yv$ (i.e., $A\mathbf{y} + \mathbf{b}$), followed by a sorting algorithm (ascending order) outputting the permutation along which the vector $A\mathbf{y} + \mathbf{b}$ is sorted -- see also Fig.~\ref{fig:Block}.
We refer to this regime as \emph{linear}.    

Characterizing the linear regime (if any) is important for several reasons.  
First, it is a natural first step to characterizing the complete solution of the problem.  
Second, in the linear regime the optimal decoder has an appealing performance from a computational complexity perspective.
The block diagram of the optimal decoder in the linear regime is shown in Fig.~\ref{fig:Block}.  The optimal decoder first computes a permutation-independent linear transformation of $\yv$ (first block in Fig.~\ref{fig:Block}), which is a polynomial in $n$ complexity task (an expression for this linear transformation is provided in Theorem~\ref{thm:MainResult} in Section~\ref{sec:MainRes}).
Next, given this linear transformation, the optimal decoder only needs to perform sorting on it (second block in Fig.~\ref{fig:Block}),
which is a task of complexity $\Oc(n \log n)$. 
Thus, in the linear regime the optimal  decoder has at most polynomial in $n$ complexity.    This performance should be compared to the brute force evaluation of the optimal test in~\eqref{region}, which has a practically prohibitive complexity of $n!$.

Currently, finding a meaningful expression for the structure of $\mathcal{R}_{\pi,K_{\mathbf{N}}}$ for all $K_{\mathbf{N}}$ seems to be a challenging task. However, some properties can be found on the structure of $\mathcal{R}_{\pi,K_{\mathbf{N}}}$ in the general case.   In particular, the following proposition, the proof of which is provided in Appendix~\ref{app:SymmDecReg}, demonstrates that the regions must have a certain symmetry.  
This property will also be useful for the characterization of the linear regime. 
\begin{proposition}\label{lem:invPoint}
Let $(\pi,\tau) \in  \mathcal{P} \times \mathcal{P}$ be the index pair that satisfies $\Hc_{\pi}=-\Hc_{\tau}$, that is $\tau_i = \pi_{n-i+1}, i \in [1:n]$, with $\pi_{i}$ and $\tau_{i}$ indicating the $i$-th element of $\pi$ and $\tau$, respectively.
Then, $\mathcal{R}_{\pi, K_{\mathbf{N}}} = - \mathcal{R}_{\tau, K_{\mathbf{N}}}$, that is for any observation $\yv \in \Rc_{\pi,K_\Nm}$ it follows that
$-\yv \in \Rc_{\tau,K_\Nm}.$
\end{proposition}

\begin{figure}
	\centering
	\includegraphics[width=345pt]{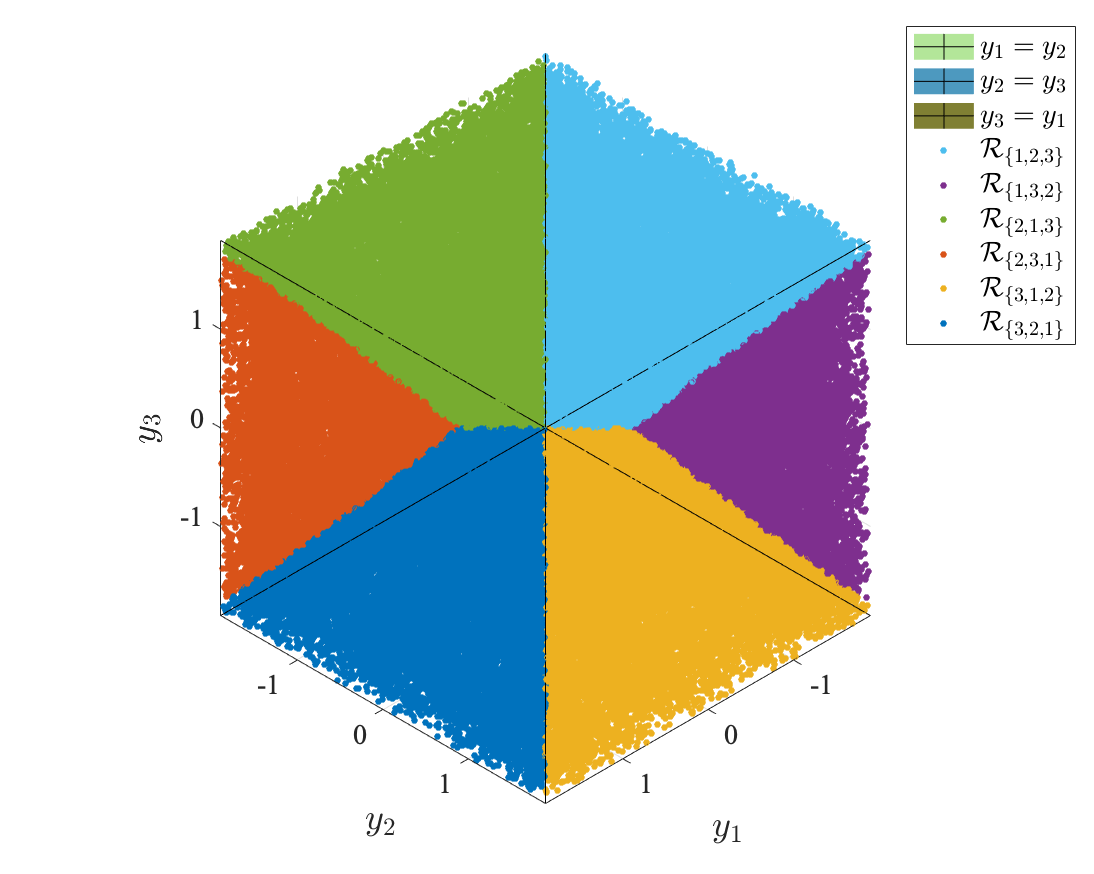}
	\caption{Monte Carlo simulation of the optimal decision regions $\Rc_{\pi,K_\Nm}, \pi \in \mathcal{P}$ where  
	$K_\Nm$ is defined in~\eqref{eq:CovEx}.  }
	\label{fig:R112}
\end{figure}

\begin{remark} \label{rem:Exch} 
We note that the result in Proposition~\ref{lem:invPoint} can be generalized beyond the Gaussian assumption on $\mathbf{X} \in \mathbb{R}^n$. In particular, it holds under the condition that $\mathbf{X} \in \mathbb{R}^n$ is an exchangeable\footnote{A sequence of random variables $U_1, U_2, \ldots, U_n$ is said to be exchangeable if,
for any permutation $(\pi_1,\pi_2,\ldots,\pi_n)$ of the indices $[1:n]$, we have that  $(U_1,U_2,\ldots,U_n) \stackrel{d}{=} (U_{\pi_1}, U_{\pi_2}, \ldots, U_{\pi_n})$,
where $\stackrel{d}{=}$ denotes equality in distribution.} random vector.  
\end{remark}
%
%
We conclude this section  by providing an example of $K_\Nm$  that puts us outside of the linear regime.  
Consider $n=3$ and the following noise covariance matrix
\begin{align}
\label{eq:CovEx}
K_\Nm=
\begin{bmatrix}
       1 & 0 & 0           \\
       0 & 1 & 0    \\
       0 & 0 & 2
\end{bmatrix}.
\end{align}
By performing  brute force comparisons in~\eqref{region},  Fig.~\ref{fig:R112} shows the structure of the optimal decision regions for the choice of  $K_\Nm$ in  \eqref{eq:CovEx}.  
We highlight that, for notational simplicity, in Fig.~\ref{fig:R112} we indicated $\Rc_{\pi,K_\Nm}$ as $\Rc_{\pi}$.
Note that the $\Hc_{\pi}$'s, which  have a cone structure (see Fig.~\ref{fig:RegN3}), cannot be a linear transformation of  the $ \Rc_{\pi,K_\Nm}$ regions in  Fig.~\ref{fig:R112}. In Section~\ref{sec:MainRes}, we will  provide a formal explanation on why the covariance matrix in \eqref{eq:CovEx} does not induce a linear regime.  
Finally, observe that as expected, in view of Proposition~\ref{lem:invPoint},  the optimal decision regions in   Fig.~\ref{fig:R112} have a point of symmetry with respect to the origin. 

\section{Main Result and Discussion} 
\label{sec:MainRes}
We here provide our main result and discuss several practically relevant implications of it.
In particular, our main result is given by the following theorem, which is proved in Section~\ref{sec:ProofMainResu}.
\begin{theorem}\label{thm:MainResult}  The following conditions are equivalent:
\begin{enumerate}
\item $\mathcal{R}_{\pi,K_{\mathbf{N}}}$ is a permutation-independent linear transformation of  $\Hc_{\pi}$; 
\item  $\mathbf{0}_n \in \bigcap_{ \pi \in \mathcal{P}}  \mathcal{R}_{\pi,K_{\mathbf{N}}}  $; 
\item The ellipsoid $\left( K_\Nm^{-1}+I_n \right)^{-\frac{1}{2}}\mathcal{B}^{n}\left(\mathbf{0}_{n},1\right)$ projected onto the hyperplane $\Wc=\{\xv \in \mathbb{R}^n:\mathbf{1}_n^T\xv=0 \}$ is an $(n-1)$-dimensional ball of radius $\gamma$ for some constant $\gamma \in (0,1)$;
\item  Let $\Qc = \left \{ Q\in \mathcal{SO}(n):  \mathbf{q}_n = \frac{1}{\sqrt{n}}\mathbf{1}_n \right \}$, where $\mathcal{SO}(n)$ is the set of $n\times n$ real-valued orthonormal matrices, and $\mathbf{q}_n$ is the $n$-th column of $Q$. Then, there exist three constants $\gamma\in (0,1)$, $a \in (0,1)$, and $v \in \mathbb{R}$ such that $v^2 < \min \{a \gamma, (1-a)(1-\gamma)\}$ and
\begin{align}
\label{eq:ConditionOnKN}
\left(K_\Nm^{-1}+I_n\right)^{-1} = Q \begin{bmatrix}
\gamma I_{n-2}  & 0_{n-2 \times 2}
\\
0_{2\times n-2} & S
\end{bmatrix} Q^T,
\end{align}
where $Q\in\Qc$ and $S = \left[\begin{smallmatrix} \gamma & v \\ v & a \end{smallmatrix} \right]$; and
\item   $\Rc_{\pi,K_\Nm} = \left(K_\Nm + I_n \right) \Hc_\pi$, for all $\pi \in \mathcal{P}$.
\end{enumerate} 
\end{theorem} 

\begin{remark}\label{remarkConditional}
 Recall that for $\Xm \sim \mathcal{N}({\bf 0}_n, I_n)$, we have that 
 \begin{align*}
 \Xm | \Ym={\bf y}  \sim \mathcal{N}\left( \mathbb{E} \left[\Xm|\Ym={\bf y}\right],  \text{\rm Var}(\Xm|\Ym=\yv)\right),
  \end{align*}
where $ \mathbb{E}\left[\Xm|\Ym={\bf y}\right]  =   (I_n+{{K}}_{\Nm} )^{-1}  {\bf y}$~\cite{kay1993fundamentals} and $\text{\rm Var}(\Xm | \Ym=\yv) =  (I_n +{K}_{\Nm}^{-1} )^{-1} ,~\forall \yv\in\mathbb{R}^n$. 
It therefore follows that condition 4) in Theorem~\ref{thm:MainResult} imposes a constraint for the conditional covariance of $\Xm$ given $\Ym$.
Moreover, recall that the conditional expectation is the optimal mean squared error estimator~\cite{kay1993fundamentals}. Therefore, the permutation-independent linear transformation in condition 5) in Theorem~\ref{thm:MainResult} is, in fact, the optimal linear estimator -- see also first block in Fig.~\ref{fig:Block}.
\end{remark}

\begin{remark}
One interesting property of condition 4) in Theorem~\ref{thm:MainResult} is the following.
Let $\Gm=\Xm|\Ym$ be the Gaussian random vector that has properties as indicated in Remark~\ref{remarkConditional}.
Then, it can be shown that $\left \{\frac{1}{i}\sum_{k=1}^i G_k - G_{i+1}\right \}, i\in[1:n-1]$ are independent.
In particular, this follows by studying $Q^T\Gm$ that has covariance given by
\begin{align}
\label{eq:equivalent4}
		Q^T  K_\Gm Q = 
		\begin{bmatrix}
		\gamma I_{n-2} & 0_{n-2\times 2} \\
		0_{2\times n-2} & S
		\end{bmatrix},
	\end{align}
where $Q \in \mathcal{Q}$ is chosen such that its element $Q_{i,j}$ in the $i$-th row and $j$-th column is 
\begin{align*}
	Q_{i,j} = 
	\begin{cases}
		(j^2+j)^{-\frac{1}{2}}, &  j\neq n, \ i \le j,  \\
		-(1+ j^{-1})^{-\frac{1}{2}}, & j\neq n, \ i=j+1, \\
		n^{-\frac{1}{2}}, & j=n, \\
0, & \text{otherwise}.
	\end{cases}
\end{align*}

\end{remark}

\begin{remark}  As discussed in Section~\ref{sec:Optimal_Region}, the computational complexity of the optimal  decoder in the linear regime is at most polynomial in $n$. It is also interesting to comment on the computational complexity of verifying whether a given $K_\Nm$ induces a linear regime.  Observe that the linearity condition in \eqref{eq:ConditionOnKN} requires to perform matrix inversion,  multiplication, and eigendecomposition. All these are polynomial in $n$ complexity tasks. Therefore, verifying if the given $K_\Nm$ satisfies~\eqref{eq:ConditionOnKN} is a polynomial in $n$ complexity task. 
\end{remark}
\begin{figure}
\begin{subfigure}{.5\textwidth}
  \centering
  \includegraphics[width=0.95\textwidth]{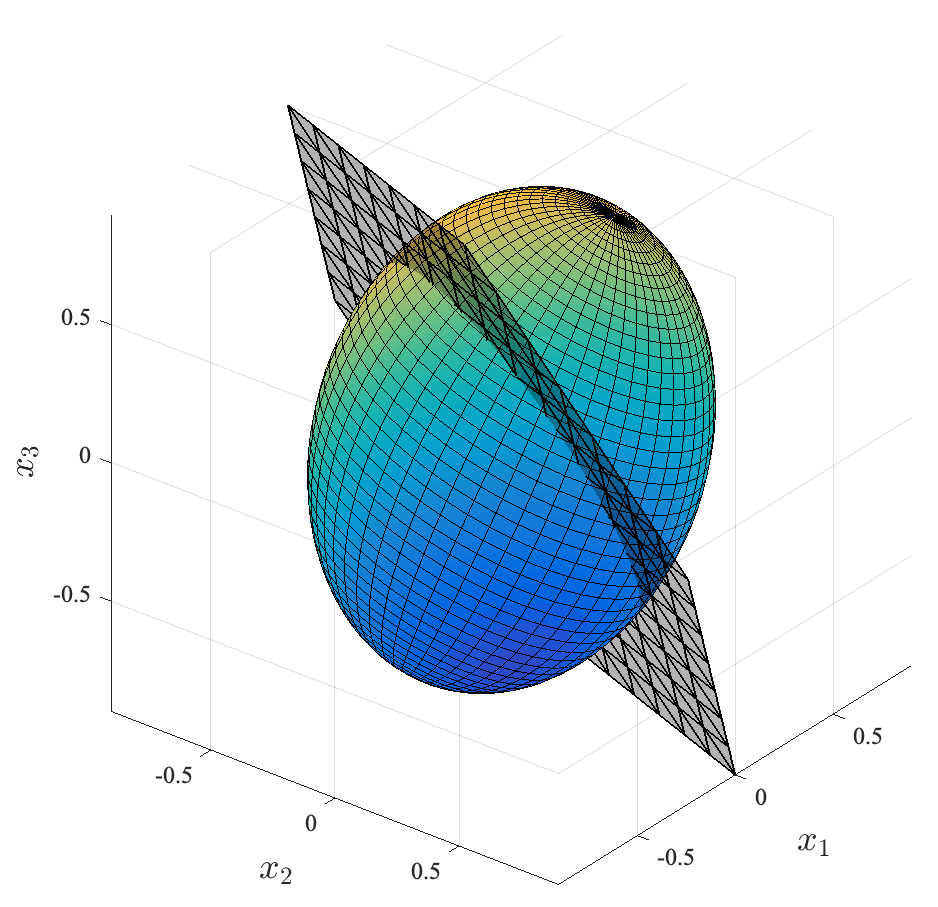}  
  \label{fig:sub-first}
\end{subfigure}
\hfill
\begin{subfigure}{.5\textwidth}
  \centering
  \includegraphics[width=0.95\textwidth]{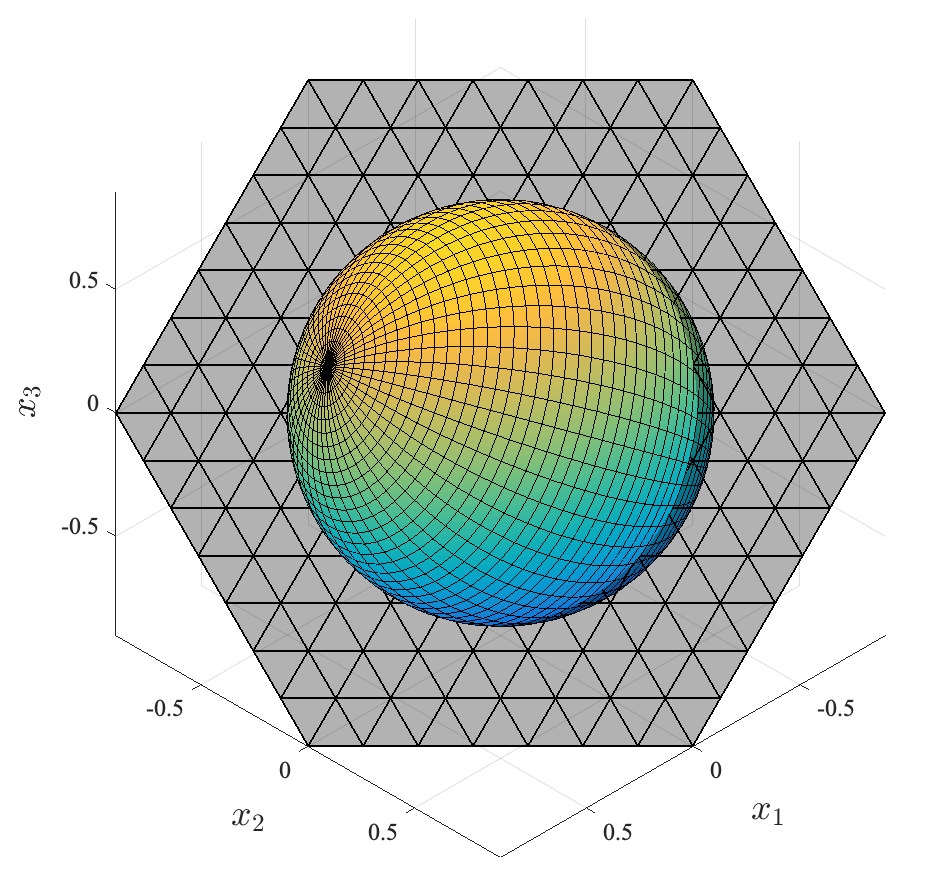}  
  \label{fig:sub-second}
\end{subfigure}
\caption{Graphical representation of the ellipsoid $\left( K_\Nm^{-1}+I_n \right)^{-\frac{1}{2}}\mathcal{B}^{n}\left(\mathbf{0}_{n},1\right)$, where $K_\Nm$ satisfies~\eqref{eq:ConditionOnKN} with parameters defined in~\eqref{eq:LinearTransf}.}
\label{fig:4ellipsoid}
\end{figure}


An example of $K_\Nm$ that induces the linear regime can be obtained by considering $n=3$ and 
\begin{align}
\label{eq:LinearTransf}
(\gamma,a,v) = \left( 0.5,0.5, 0.2 \right)
\end{align}
in~\eqref{eq:ConditionOnKN}. By taking the eigendecomposition of this $K_\Nm$, it can be verified that it has three distinct eigenvalues given by
$\lambda_1 = 1$, $\lambda_2 = 3/7$ and $\lambda_3 = 7/3$.
The corresponding ellipsoid $\left( K_\Nm^{-1}+I_n \right)^{-\frac{1}{2}}\mathcal{B}^{n}\left(\mathbf{0}_{n},1\right)$ has three distinct radii and it is shown in Fig.~\ref{fig:4ellipsoid} (left). The projection of this ellipsoid onto $\Wc=\{\xv \in \mathbb{R}^3:\mathbf{1}_3^T\xv=0 \}$ is equal to a $2$-dimensional ball of radius $\gamma=1/2$ as also illustrated in Fig.~\ref{fig:4ellipsoid} (right).
\begin{figure}
	\centering
	\includegraphics[width=300pt]{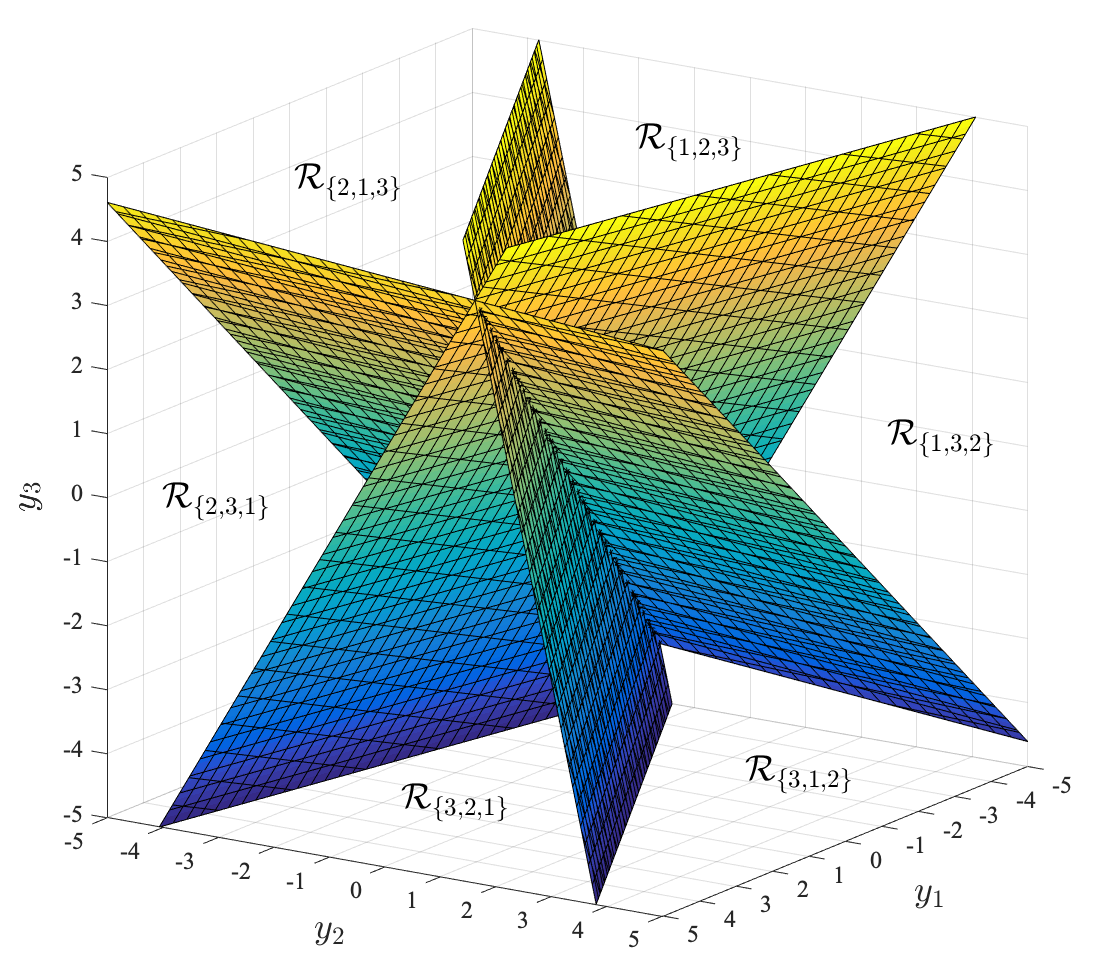}
	\caption{Optimal decision regions of the $K_\Nm$ that satisfies~\eqref{eq:ConditionOnKN} with parameters defined in~\eqref{eq:LinearTransf}.}
	\label{fig:R}
\end{figure}
Fig.~\ref{fig:R} shows that the corresponding optimal decision regions $\Rc_{\pi,K_\Nm}, \pi \in \mathcal{P}$, are indeed obtained as a permutation-independent linear transformation of the corresponding hypothesis regions in Fig.~\ref{fig:RegN3}, namely as $\Rc_{\pi,K_\Nm} = \left(K_\Nm + I_3 \right) \Hc_\pi$. We highlight that, for notational simplicity, in Fig.~\ref{fig:R} we indicated $\Rc_{\pi,K_\Nm}$ as $\Rc_{\pi}$.

%
%
%

\subsection{Sufficient and Necessary Conditions on the Spectrum and on the Eigenvectors of $K_\Nm$}
We here provide necessary and sufficient conditions on the spectrum of $K_\Nm$, i.e., on the set of its eigenvalues, as well as on its eigenvectors that need to be satisfied for~\eqref{eq:ConditionOnKN} to hold.
In particular, we have the next proposition, the proof of which can be found in Appendix~\ref{app:EigenvaluesEigenvectors}.

\begin{proposition}
\label{prop:Spectrum}
A $K_\Nm$ satisfies the condition in~\eqref{eq:ConditionOnKN} if and only if it has eigenvalues $\lambda_i,~i\in[1:n]$ and eigenvectors $\nuv_i,~i\in[1:n]$ that are in either one of the two forms below:
\begin{itemize}
\item Case 1: All the $n$ eigenvalues are the same; we have
\begin{align}
\label{eq:EigenFiniso}
\lambda_i = \frac{\gamma}{1-\gamma},
\qquad
\nuv_i = \tv_i,
\end{align}
where $\{ \tv_i,~i\in[1:n] \}$ is any set of orthogonal vectors in $\mathbb{R}^n$, and $\gamma \in (0,1)$;
\item Case 2: At least two eigenvalues are different; we have
\begin{subequations}\label{eq:EigenvaluesFinNon}
\begin{align}
\label{eq:EigenvaluesFin}
\lambda_i & =
\left \{
\begin{array}{ll}
\frac{\gamma}{1-\gamma}, & i \in [1: n-2],
\\
\frac{a+\gamma + \sqrt{(a-\gamma)^2 + 4 {v}^2}}{2-a-\gamma - \sqrt{(a-\gamma)^2 + 4{v}^2}}, & i =n-1,
\\
\frac{a+\gamma - \sqrt{(a-\gamma)^2 + 4 {v}^2}}{2-a-\gamma + \sqrt{(a-\gamma)^2 + 4{v}^2}}, & i =n,
\end{array}
\right .
\\
	\nuv_i & = 
	\begin{cases}
		\qv_i, & i\in[1:n-2], \\
		\left (v+a-\frac{\lambda_i}{1+\lambda_i} \right ) \mathbf{q}_{n-1} + \left (v+\gamma-\frac{\lambda_i}{1+\lambda_i} \right)\qv_n, &  i \in [n-1:n], 
	\end{cases}
\label{eq:EigenvectorsFin}
\end{align}
\end{subequations}
where $\gamma \in (0,1)$, $a \in (0,1)$, $v \in \mathbb{R}$ satisfying $v^2 < \min \{a \gamma, (1-a)(1-\gamma)\}$, and $\mathbf{q}_i, i \in [1:n]$ is the $i$-th column of $Q \in \mathcal{Q}$.
\end{itemize}
%
\end{proposition}

\begin{remark}
Proposition~\ref{prop:Spectrum} provides necessary and sufficient conditions for $K_\Nm$ to satisfy Theorem~\ref{thm:MainResult} in terms of its eigenvalues and eigenvectors. Specifically, Proposition~\ref{prop:Spectrum} shows that a $K_\Nm$ that satisfies~\eqref{eq:ConditionOnKN}  has at most three distinct eigenvalues.

%
\end{remark}

\subsection{Case of $n=2$ is Special}
It is interesting to note that in the  case of $n=2$ the condition in~\eqref{eq:ConditionOnKN} is not  restrictive, i.e., all covariance matrices satisfy~\eqref{eq:ConditionOnKN}. 
To put it in other words, for $n=2$ the linear regime is the only regime, and Theorem~\ref{thm:MainResult} gives a complete characterization of the permutation recovery problem.  

One intuitive explanation why this follows is given by condition 3) in Theorem~\ref{thm:MainResult} which requires that the projection of an $n$-dimensional ellipsoid onto the hyperplane $\Wc$ is an $(n-1)$-dimensional ball. When $n=2$, this corresponds to projecting an ellipse onto a line. The result of this operation is a segment, which is indeed a $1$-dimensional ball. Therefore, for the case of $n=2$ any $K_\Nm$ satisfies~\eqref{eq:ConditionOnKN}. We next prove this formally using condition 4) in Theorem~\ref{thm:MainResult}.

\begin{proposition}  
\label{prop:n2}
Let $n=2$.  Then, every positive definite covariance matrix $K_{\mathbf{N}}$ satisfies  \eqref{eq:ConditionOnKN}. 
\end{proposition} 
\begin{IEEEproof}
For $n=2$ and any positive definite symmetric $K_\Nm$, the left-had side of~\eqref{eq:ConditionOnKN} can be represented by the triple $(w,q,z)$ as
\begin{align}\label{eq:leftCond}
(K_\Nm^{-1}+I_n)^{-1} = 
\begin{bmatrix}
w & q \\
q & z
\end{bmatrix},
\end{align} 
where $w>0$, $z>0$, and $w z > q^2$. Note also that the eigenvalues of the left-hand side of~\eqref{eq:leftCond} are smaller than one, and hence the triple $(w,q,z)$ has also to satisfy this constraint. 
Hence, we would need to find a triple $(a,\gamma,v)$ such that
\begin{align}\label{eq:n2cond}
\begin{bmatrix}
w & q \\
q & z
\end{bmatrix}
& = 
Q
\begin{bmatrix}
\gamma & v \\
v & a
\end{bmatrix}
Q^T,
\end{align} where the orthonormal matrix $Q \in \mathcal{Q}$ can be chosen as
\begin{align}
Q = 
\frac{1}{\sqrt{2}} \begin{bmatrix}
-1 & 1 \\
1 & 1
\end{bmatrix}.
\end{align}
It is not difficult to see that the triple $(a,\gamma,v)$ such that
\begin{align*}
&a = \frac{w+z+2q}{2}, \quad \gamma = \frac{w+z-2q}{2}, \quad v = \frac{z-w}{2},
\end{align*}
satisfies all the constraints in condition 4) of Theorem~\ref{thm:MainResult}. This concludes the proof of Proposition~\ref{prop:n2}.
\end{IEEEproof}  

\subsection{For $n>2$ Memoryless Noise Can Only be Isotropic} 
We here focus on the case $n>2$, and we prove that if the noise is memoryless, i.e., $K_\Nm$ is a diagonal matrix, then all its diagonal elements has to be equal to ensure that~\eqref{eq:ConditionOnKN} is satisfied, i.e., the noise has to be isotropic. 
We note that this result justifies the fact that the $K_\Nm$ defined in~\eqref{eq:CovEx} puts us outside of the linear regime (see Fig.~\ref{fig:R112}).
We also highlight that such a restriction does not apply for the case $n=2$ since, as we have shown in Proposition~\ref{prop:n2}, for this case any $K_\Nm$ satisfies~\eqref{eq:ConditionOnKN}. 
\begin{proposition}\label{prop:diagonal}
Consider $n>2$ and let $K_\Nm$ be a diagonal positive definite matrix. Then, $K_\Nm$ satisfies~\eqref{eq:ConditionOnKN} if and only if
\begin{align}
\label{eq:IsotNoise}
K_\Nm = \frac{\gamma}{1-\gamma} I_n,
\end{align}
for some $\gamma \in (0,1)$.
\end{proposition}

\begin{IEEEproof}
Let $K_\Nm\in\mathbb{R}^{n\times n},n>2$ be a diagonal matrix with $\sigma_i^2,~i\in[1:n]$ on its diagonal entries. We start by observing that if $K_\Nm$ is isotropic (i.e., $K_\Nm = c I_n$ for any constant $c>0$), then it has eigenvalues $\lambda_i$ and eigenvectors $\nuv_i$ as in~\eqref{eq:EigenFiniso} in Proposition~\ref{prop:Spectrum}. Thus, if $K_\Nm$ is isotropic, then it satisfies the condition in~\eqref{eq:ConditionOnKN}.

We now show that any diagonal positive definite $K_\Nm$ has to be of the form as in~\eqref{eq:IsotNoise} to satisfy~\eqref{eq:ConditionOnKN}.
Towards this end, assume that $K_\Nm$ is non-isotropic. For $i\in[1:n]$, since $K_\Nm$ is a diagonal matrix, it has eigenvalues $\lambda_i$ and eigenvectors $\nuv_i$ given by
\begin{align}
\label{eq:nonisoee}
	\lambda_i  = \sigma_i^2, \qquad \nuv_i  = \ev_i,
\end{align}
where $\ev_i\in\mathbb{R}^n$ denotes an $n$-dimensional vector of all-zeros except a non-zero element in the $i$-th position. 
However, from Proposition~\ref{prop:Spectrum}, we know that there exists $i \in [1:n]$ for which $\nuv_{i} = \frac{\gamma-a}{\sqrt{n}}\mathbf{1}_n$ (since $v=0$), and hence a $K_\Nm$ that is diagonal, but non-isotropic does not satisfy the condition in~\eqref{eq:ConditionOnKN}.
This concludes the proof of Proposition~\ref{prop:diagonal}.
\end{IEEEproof}

\subsection{On the Probability of Error} 
Although finding the probability of error is not the main objective of this paper, we make a few comments about it.  Specifically, the structure of the optimal decision regions in Theorem~\ref{thm:MainResult} can now be utilized to provide the following geometric characterization of the error probability, the proof of which can be found in Appendix~\ref{app:ErrorProb}.
\begin{proposition}\label{prop:ProbError} Let $K_{\mathbf{N}}$ satisfy the conditions in Theorem~\ref{thm:MainResult}. Then, the error probability is given~by  
\begin{subequations}
\label{eq:Pe}
\begin{align} 
P_e= 1- n! \frac{   \mathrm{Vol}^{2n}\left(\mathcal{C}_{\mathcal{H}_\pi}\cap A \mathcal{B}^{2n}\left(\mathbf{0}_{2n},1\right)\right)}{  \mathrm{det} \left ( K_{\mathbf{N}}^{ \frac{1}{2} } \right )  \mathrm{Vol}^{2n}\left(\mathcal{B}^{2n}\left(\mathbf{0}_{2n},1\right)\right)}, 
\end{align} 
where
\begin{align}
A=\begin{bmatrix} {I}_n& {0}_{n\times n}\\ {I}_n&  K_{\mathbf{N}}^{  \frac{1}{2} } \end{bmatrix}, \quad
\mathcal{C}_{\mathcal{H}_{\pi}}=  \mathcal{H}_{\pi} \times  (K_{\mathbf{N}}+I_n) \mathcal{H}_{\pi},
\end{align} 
\end{subequations}
and where $\pi \in \mathcal{P}$
can be chosen arbitrarily. 
\end{proposition} 

The result in Proposition~\ref{prop:ProbError} can now be used to derive various upper and lower bounds on the probability of error, and hence find {\em impossibility} results, i.e., properties on the noise covariance matrix $K_{\mathbf{N}}$ for which reasonable recovery is not possible.  This is an interesting direction, which we leave for future work. 
The interested reader is referred to~\cite{jeong20}  for a preliminary work in this direction that shows that the problem considered in this work is noise limited.


\subsection{Discussion on Possible Extensions} 
We here discuss a few possible future directions and extensions. 
Perhaps one of the most natural next directions is to look beyond the linear regime.  For example, it would be interesting to understand whether the optimal decoder always has a reasonable closed-form characterization.  In particular, Proposition~\ref{lem:invPoint} and the simulation results in Fig.~\ref{fig:R112} suggest that the optimal decision regions have a symmetrical polyhedral structure, and it would be interesting to see if the general structure of the optimal decision regions can be characterized.  The possibility that such a general characterization exists stems from the following characterization of the optimal decoder: given an observation ${\bf y}$ 
\begin{align}
\pi^\star= \arg \max_{\pi \in \mathcal{P}}  \Pr[ {\bf X} \in \mathcal{H}_\pi | {\bf Y}={\bf y}  ]\!=\! \arg \max_{\pi \in \mathcal{P}}  \Pr[  (I_n+K_{\mathbf{N}}) ^{-1} {\bf y}  +(I_n+K_{\mathbf{N}}^{-1}) ^{-\frac{1}{2}}{\bf Z} \in \mathcal{H}_\pi   ], \label{eq:Optimal_Decoder}
\end{align} 
where ${\bf Z}$ is a standard Gaussian random vector.  The proof of the second equality in \eqref{eq:Optimal_Decoder} follows from the fact that ${\bf X}$ given ${\bf Y}$ is Gaussian; see Remark~\ref{remarkConditional} for more details.   

It would also be interesting to study the probability of error for the linear decoder proposed in this work and compare it with the probability of error of the optimal decoder in the regimes not covered by Theorem~\ref{thm:MainResult}.
Recall that the optimal decoder in the linear regime consists of the optimal linear estimator combined with a sorting operation (see Remark~\ref{remarkConditional} and Fig.~\ref{fig:Block}). This decoder is very attractive as it is relatively easy to implement in practice. In particular, it is reasonable to suspect that there exists a large set of noise covariance matrices for which such a decoder will perform relatively well.   

Another interesting direction is to consider whether the results of this paper can be generalized beyond the assumption that ${\bf X} \in \mathbb{R}^n$ is Gaussian.  One attractive direction to consider is the case when ${\bf X} \in \mathbb{R}^n$ is  exchangeable. The assumption of exchangeability still allows to use the symmetry argument, and in particular, Proposition~\ref{lem:invPoint} holds under this assumption (see Remark~\ref{rem:Exch}).  Furthermore,  let ${\bf X}_{\bf y} \in \mathbb{R}^n$  be the random variable distributed according to $f_{{\bf X}|{\bf Y}}( \cdot | {\bf y})$; then, from Proposition~\ref{lem:invPoint} it follows that the linear regime is optimal if and only if there exists a constant $c \in (0,1)$ such that
\begin{align}
\Pr[{\bf X}_0 \in \mathcal{H}_\pi] =c, \forall \pi \in \mathcal{P}.  \label{eq:Condition_for_conditional}
\end{align}
 In our preliminary work in~\cite{jeong20}, we have shown the optimality of the linear regime when the noise is  isotropic.
Thus, an interesting future direction would consist of identifying the family of the noise covariance matrices for which~\eqref{eq:Condition_for_conditional} holds when ${\bf X} \in \mathbb{R}^n$ is  exchangeable, but not necessarily Gaussian.

%

\section{Proof of Theorem~\ref{thm:MainResult}} 
\label{sec:ProofMainResu}
In this section, we prove the results in Theorem~\ref{thm:MainResult}. 
In particular, the proof follows the next sequence of implications  
\begin{align*}
1) \Rightarrow 2) \Leftrightarrow 3)  \Leftrightarrow 4) \Rightarrow 5)   \Rightarrow 1),
\end{align*} 
which are next analyzed in different subsections.
Note that the implication $5)   \Rightarrow 1)$ follows immediately.

\subsection{Proof of the Implication  $1) \Rightarrow 2)$} 

We here prove that $1) \Rightarrow 2)$, i.e., the fact that $\mathcal{R}_{\pi,K_{\mathbf{N}}}$ is a permutation-independent linear transformation of  $\Hc_{\pi}$ implies that $\mathbf{0}_n \in \bigcap_{ \pi \in \mathcal{P}}  \mathcal{R}_{\pi,K_{\mathbf{N}}}$.
Towards this end, we prove the following lemma by leveraging the symmetry condition proved in Proposition~\ref{lem:invPoint}.
\begin{lemma}\label{lem:nonlinear}
Suppose that 
\begin{align}
\Rc_{\pi,K_\Nm}=A\Hc_\pi + \bv,~ \forall \pi\in\Pc,
\end{align} 
where $A$ is an $n\times n$ matrix, and $\bv$ is an $n$-dimensional column vector. 
Then,   $\mathbf{0}_n \in \bigcap_{ \pi \in \mathcal{P}}  \mathcal{R}_{\pi,K_{\mathbf{N}}}$.   Moreover, $\bv$ must be of the form $\bv= t A \mathbf{1}_n$ for some $t\in \mathbb{R}$. 
\end{lemma}
\begin{IEEEproof}
Let $\mathcal{L}_{\Hc}= \left\{  \xv \in \mathbb{R}^n:\xv\in \bigcap_{\pi\in\Pc}\Hc_\pi \right \}$ be the set of points that belong to the intersection of $\Hc_\pi,~\forall \pi\in\Pc$. Note that this set of points forms a  line in $\mathbb{R}^n$, which is given~by
\begin{align}
\label{eq:CH}
\mathcal{L}_{\Hc}=\left\{  \xv \in \mathbb{R}^n:\xv=\kappa \mathbf{1}_n,{~\kappa \in\mathbb{R}}\right\}.
\end{align} 
Similarly, let $\mathcal{L}_\Rc=\left\{  \xv \in \mathbb{R}^n: \xv\in\bigcap_{\pi\in\Pc}\Rc_{\pi,K_\Nm} \right\}$ be the set of points that belong to  the intersection of $\Rc_{\pi,K_\Nm},~\forall \pi\in\Pc$. Note that this set is non-empty.  
From the assumption in Lemma~\ref{lem:nonlinear}, we have that  $\mathcal{L}_\Rc = A\mathcal{L}_\Hc+ \bv$. 
Thus,  $\mathcal{L}_\Rc$ is also a line in $\mathbb{R}^n$ defined as
\begin{align}
\mathcal{L}_\Rc=\left\{ \xv \in \mathbb{R}^n:\xv=\kappa  A \mathbf{1}_n+ \bv,{~\kappa \in\mathbb{R}}\right\}.
\end{align}
Now let $\mathbf{0}_n \neq \tilde{\yv} \in \mathcal{L}_\Rc$.  Then, by Proposition~\ref{lem:invPoint} if $ \tilde{\yv} \in \mathcal{L}_\Rc$, we have that $-  \tilde{\yv} \in \mathcal{L}_\Rc $.  Since $\mathcal{L}_\Rc
$  is a line that contains both $-  \tilde{\yv} $ and $  \tilde{\yv}$,  it must contain also $\mathbf{0}_n$.  
Finally, observe that the only $\bv$ that is allowed (i.e., that ensures that the line contains both $-  \tilde{\yv} $ and $  \tilde{\yv}$) is of the form $\bv= t A \mathbf{1}_n$ for some $t\in \mathbb{R}$. This concludes the proof of Lemma~\ref{lem:nonlinear}.
\end{IEEEproof}
Note that the fact that the shift vector $\bv$ in Lemma~\ref{lem:nonlinear} is of the form $\bv= t A \mathbf{1}_n$, for some $t\in \mathbb{R}$, implies that
\begin{align}
\mathcal{L}_\Rc &=\left\{ \xv \in \mathbb{R}^n:\xv=\kappa  A \mathbf{1}_n+ \bv,{~\kappa \in\mathbb{R}}\right\} \nonumber
\\& =\left\{ \xv \in \mathbb{R}^n:\xv=(\kappa+t)  A \mathbf{1}_n,{~\kappa,t \in\mathbb{R}}\right\} =A \mathcal{L}_\Hc,
\end{align} 
and
\begin{align}
\Rc_{\pi,K_\Nm}=A\Hc_\pi + \bv= A (\Hc_\pi + t  \mathbf{1}_n)=  A \Hc_\pi. 
\end{align}
In other words, such a choice of $\bv$ does not effect the shape of the decision regions. 

\subsection{Proof of the Implication  $2) \Leftrightarrow 3)$} 
We here prove that  $2) \Leftrightarrow 3)$, i.e., the fact that the ellipsoid $\left( K_\Nm^{-1}+I_n \right)^{-\frac{1}{2}}\mathcal{B}^{n}\left(\mathbf{0}_{n},1\right)$ projected onto the hyperplane $\Wc=\{\xv \in \mathbb{R}^n:\mathbf{1}_n^T\xv=0 \}$ is an $(n-1)$-dimensional ball of radius $\gamma$ for some $\gamma \in (0,1)$ implies that $\mathbf{0}_n \in \bigcap_{ \pi \in \mathcal{P}}  \mathcal{R}_{\pi,K_{\mathbf{N}}}  $, and vice versa.

In particular, the  proofs $2) \Leftarrow 3)$ and $2) \Rightarrow 3)$ will leverage a symmetrization method known as Steiner symmetrization~\cite{gruber2007convex}, which we next formally define.
\begin{definition} 
\label{def:ST}
Let $\mathcal{S}$ be a bounded set in $\mathbb{R}^n$, and $\Wc$ be an $(n-1)$-dimensional vector subspace of $\mathbb{R}^n$. The Steiner symmetrization of $\mathcal{S}$ with respect to $\Wc$ is the operation that associates the set $\mathrm{st}_\Wc(\mathcal{S})$ in $\mathbb{R}^n$ to the set $\mathcal{S}$ such that, for each straight line $\ell$ perpendicular to $\Wc$, we have that $\ell \cap \mathcal \mathrm{st}_\Wc(\mathcal{S})$ is either a closed line segment with center in $\Wc$ or is empty. 
Moreover, the two following conditions need to be satisfied
\begin{subequations}
\begin{align}
\mathsf{length} \left ( \ell \cap \mathcal{S} \right ) = \mathsf{length}  \left ( \ell \cap \mathrm{st}_\Wc(\mathcal{S}) \right ),
\end{align}
and
\begin{align}
\ell \cap \mathrm{st}_\Wc(\mathcal{S}) = \varnothing \quad \text{if and only if} \quad \ell \cap \mathcal{S} = \varnothing. 
\end{align}
\end{subequations}
\end{definition}
\begin{figure}
	\centering
	\includegraphics[width=290pt]{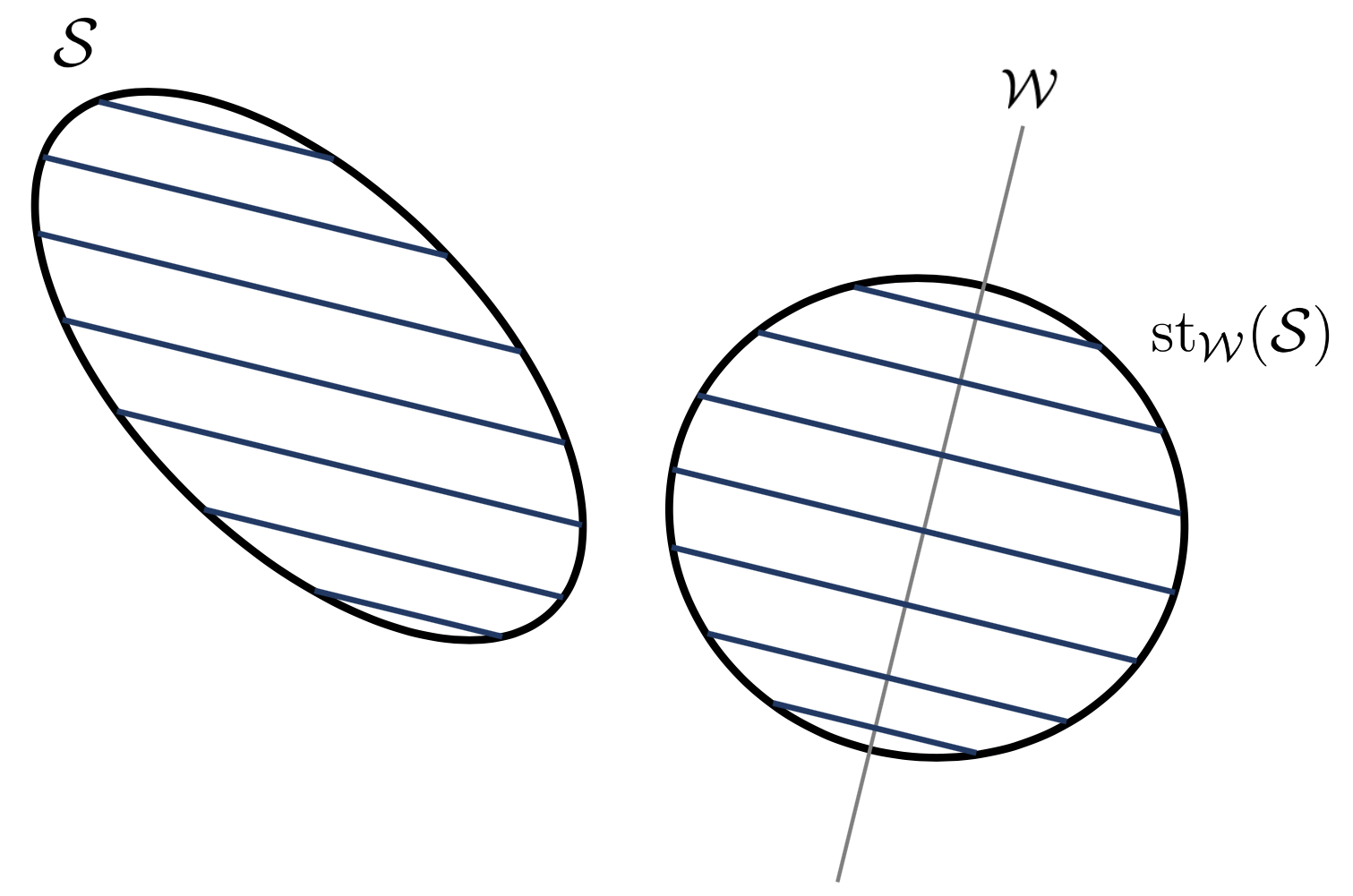}
	\caption{Steiner symmetrization.}
	\label{fig:ST}
\end{figure}
Fig.~\ref{fig:ST} illustrates the application of Steiner symmetrization on the set $\mathcal{S}$ with respect to the line $\Wc$. We now provide some properties of Steiner symmetrization that will be useful in the upcoming proofs.
\begin{proposition}
\label{prop:STprop}
The Steiner symmetrization $\mathrm{st}_\Wc(\mathcal{S})$ of the set $\mathcal{S}$ with respect to $\Wc$ satisfies the following  properties:
\begin{itemize}
\item Steiner symmetrization preserves convexity. Moreover, Steiner symmetrization transforms ellipsoids into ellipsoids~\cite{bourgain1989estimates}.
\item Steiner symmetrization preserves the volume, i.e., $\mathrm{Vol}^n\left( \mathcal{S} \right) = \mathrm{Vol}^n\left( \mathrm{st}_\Wc(\mathcal{S})\right)$~\cite{gruber2007convex}.
%
\item Steiner symmetrization preserves the orthogonal projection onto $\Wc$, i.e., ${\rm Proj}_{\Wc}(\Sc) = {\rm Proj}_{\Wc}({\rm st}_{\Wc}(\Sc))$, where ${\rm Proj_\Wc(\Ac)}$ denotes the orthogonal projection of the set $\Ac$ onto~$\Wc$~\cite{klain2012steiner}.
\end{itemize}
\end{proposition}
Another result that we will leverage to prove $2) \Leftrightarrow 3)$ is provided by the following lemma, the proof of which can be found in Appendix~\ref{app:PrNinHpi}. 
\begin{lemma}\label{lem:GaussianVolumue}
Let $\mathbf{U} \sim\Nc(\mathbf{0}_n,K_{\mathbf{U}})$, where $K_{\mathbf{U}}$ is positive definite. 
Then, 
\begin{align}
\Pr( \mathbf{U}  \in\Hc_\pi) 
& =  \frac{\left |{\rm det} \left (K_{\mathbf{U}}^{-\frac{1}{2}} \right ) \right | \mathrm{Vol}^n\left( \Hc_\pi \cap K_{\mathbf{U}}^{\frac{1}{2}} \mathcal{B}^{n}\left(\mathbf{0}_{n},1\right)\right)}{\mathrm{Vol}^n\left(\mathcal{B}^{n}\left(\mathbf{0}_{n},1\right)\right)}.
\end{align}
\end{lemma}
We are now ready to prove $2) \Leftrightarrow 3)$, the proof of which consists of two parts. The first part is provided in the next lemma, which leverages the observation in Remark~\ref{remarkConditional} and is proved in Appendix~\ref{app:VolumeConsta}.
%
\begin{lemma}\label{lem:conditionY0}
 $\mathbf{0}_n\in \bigcap_{\pi \in \mathcal{P}}\Rc_{\pi,K_\Nm} $
if and only if there exists a constant $\eta >0$ such that
\begin{align}
\mathrm{Vol}^n\left( \Hc_\pi \cap \left( K_\Nm^{-1}+I_n \right)^{-\frac{1}{2}}\mathcal{B}^{n}\left(\mathbf{0}_{n},1\right)\right) \label{eq:Vol0}
& = 	\eta,~\forall \pi \in \Pc.
\end{align} 
\end{lemma}
The second part of the proof $2) \Leftrightarrow 3)$ is given by the next lemma, which characterizes the solution of~\eqref{eq:Vol0} in terms of $K_\Nm$ and relies on the Steiner symmetrization technique. 
%

%
\begin{lemma}\thmlabel{BoundaryAt0}
A $K_\Nm $ is a solution for~\eqref{eq:Vol0} if and only if  there exists 
a constant $\gamma \in (0,1)$ such that the ellipsoid $\left( K_\Nm^{-1}+I_n \right)^{-\frac{1}{2}}\mathcal{B}^{n}\left(\mathbf{0}_{n},1\right)$ projected onto the hyperplane $\Wc=\{\xv \in \mathbb{R}^n:\mathbf{1}_n^T\xv=0 \}$ is an $(n-1)$-dimensional ball of radius $\gamma$.
\end{lemma}
\begin{IEEEproof}
Let $\mathcal{L}_{\Hc}= \left\{ \xv \in \mathbb{R}^n:\xv\in \bigcap_{\pi\in\Pc}\Hc_\pi \right \}$ be the set of points that belong to the intersection of $\Hc_\pi,~\forall \pi\in\Pc$. From~\eqref{eq:CH}, we have that
\begin{align}
\label{eq:BoundHypReg}
\mathcal{L}_{\Hc}=\left\{\xv \in \mathbb{R}^n:\xv=\kappa \mathbf{1}_n,~\kappa \in\mathbb{R}\right\}, 
\end{align}
which is a line in $\mathbb{R}^n$. 
From Lemma~\ref{lem:conditionY0}, we have that $\yv=\mathbf{0}_n$ is a boundary point for all the optimal decision regions, i.e., $\mathbf{0}_n\in  \bigcap_{\pi \in \mathcal{P}}\Rc_{\pi,K_\Nm} $, if and only if 
\begin{align}\label{eq:Vol0app}
\mathrm{Vol}^n\left( \Hc_\pi \cap \left( K_\Nm^{-1}+I_n \right)^{-\frac{1}{2}}\mathcal{B}^{n}\left(\mathbf{0}_{n},1\right)\right)
& = 	\eta,~\forall \pi \in \Pc,
\end{align}
for  some $\eta >0$.
In particular, with reference to~\eqref{eq:Vol0app},
$\Hc_\pi$ is an $n$-dimensional cone, and $\left(K_\Nm^{-1}+I_n \right)^{-\frac{1}{2}}\Bc^n(\mathbf{0}_n,1)$ is an $n$-dimensional ellipsoid centered at $\mathbf{0}_n$. We also highlight that $\Hc_\pi,~\forall\pi$ are all open sets along the direction $\mathcal{L}_{\Hc}$, i.e., for any $\pi\in\Pc$ and $\kappa\in\mathbb{R}$, if $\tilde{\xv}\in\Hc_\pi$, then $\tilde{\xv} + \kappa \mathbf{1}_n \in \Hc_\pi$.

For ease of geometrical representation, we now apply Steiner symmetrization (see Definition~\ref{def:ST}) on the ellipsoid $\left(K_\Nm^{-1}+I_n \right)^{-\frac{1}{2}}\Bc^n(\mathbf{0}_n,1)$. In particular, with reference to Definition~\ref{def:ST}, we consider the Steiner symmetrization with respect to the hyperplane
\begin{align}
\label{eq:HypST}
\Wc=\{\xv \in \mathbb{R}^n:\mathbf{1}_n^T\xv=0 \},
\end{align}
which is perpendicular to the line $\mathcal{L}_{\Hc}$ in~\eqref{eq:BoundHypReg}. 
Note that $\Wc$ is an $(n-1)$-dimensional vector subspace of $\mathbb{R}^n$.
By applying Steiner symmetrization on the ellipsoid $\left(K_\Nm^{-1}+I_n \right)^{-\frac{1}{2}}\Bc^n(\mathbf{0}_n,1)$ with respect to $\Wc$ in~\eqref{eq:HypST}, we obtain a new ellipsoid $\Ec^n$ (see Proposition~\ref{prop:STprop}) given~by
\begin{align}
\label{eq:NewEllips}
\Ec^n=\mathrm{st}_\Wc\left(\left(K_\Nm^{-1}+I_n \right)^{-\frac{1}{2}}\Bc^n(\mathbf{0}_n,1)\right),
\end{align}
which has the same volume of the original ellipsoid (see Proposition~\ref{prop:STprop}), namely
\begin{align*}
\mathrm{Vol}^n\left( \left(K_\Nm^{-1}+I_n \right)^{-\frac{1}{2}}\Bc^n(\mathbf{0}_n,1) \right ) = \mathrm{Vol}^n\left(\Ec^n \right ).
\end{align*}
It is also worth noting that $\Ec^n$ is centered at $\mathbf{0}_n$, it has $\mathcal{L}_\Hc$ in~\eqref{eq:BoundHypReg} as an axis, and it is symmetric with respect to $\Wc$. These properties, together with the fact that $\Hc_\pi$'s with $\pi \in \mathcal{P}$ are all open sets along the direction $\mathcal{L}_{\Hc}$, imply that
\begin{align}\label{eq:SteinerVol}
& \mathrm{Vol}^n\left( \Hc_\pi \cap \left( K_\Nm^{-1}+I_n \right)^{-\frac{1}{2}}\mathcal{B}^{n}\left(\mathbf{0}_{n},1\right)\right) = \mathrm{Vol}^n\left(  \Hc_\pi \cap \Ec^n \right).
\end{align}
A graphical representation of the procedure explained above is provided in Fig.~\ref{fig:STellips} for the $2$-dimensional case. 
\begin{figure}
	\centering
	\input{STexample_V4.tex}
	\captionsetup{singlelinecheck=off}
	\caption{Steiner symmetrization of the ellipsoid $\mathcal{K}=\left(K_\Nm^{-1}+I_2 \right)^{-\frac{1}{2}}\Bc^2(\mathbf{0}_2,1)$ with respect to $\Wc$ in~\eqref{eq:HypST} where $K_{\Nm}=\left [\protect\begin{smallmatrix} \frac{1}{3}&0\\ 0&4\protect\end{smallmatrix}\right ]$.    }
	\label{fig:STellips}
\end{figure}
From the analysis above, it therefore follows that the problem of finding the family of $K_\Nm$'s that satisfies~\eqref{eq:Vol0app} is equivalent to finding the family of $K_\Nm$'s such that there exists a constant $\eta >0$ for which
\begin{align}
\label{eq:ConstVolST}
\mathrm{Vol}^n\left(  \Hc_\pi \cap \Ec^n \right) = \eta, ~\forall \pi \in \Pc.
\end{align}
We now leverage the following lemma, the proof of which can be found in Appendix~\ref{app:EllipsEqualRadii}, which provides sufficient and necessary conditions for~\eqref{eq:ConstVolST} to hold.
\begin{lemma}\label{n-1ball}
Let $\Ec^n$ be an $n$-dimensional ellipsoid centered at the origin and having one axis of the type $\nuv = \frac{1}{\sqrt{n}} \mathbf{1}_n$.
Then,  there exists $\eta >0$, such that 
\begin{align}
{\rm Vol}^n\left(\Hc_\pi \cap \Ec^n \right)=\eta,~\forall\pi\in\Pc,
\end{align}
if and only if $\Ec^n$ has equal radii for all axes except possibly the axis $\nuv$.
\end{lemma}
The result in Lemma~\ref{n-1ball} says that, in order for~\eqref{eq:ConstVolST} to hold, the ellipsoid $\Ec^n$ has to have a special structure, namely it has to have equal radii for all axes except possibly the axis $\mathcal{L}_\Hc$ in~\eqref{eq:BoundHypReg}.
Mathematically, this special structure of the ellipsoid $\Ec^n$ can be represented as
\begin{align}\label{eq:intersectBall}
\Ec^n \cap \Wc = \Bc^{n-1}(\mathbf{0}_n,\gamma),
\end{align} where $\gamma\in (0,1)$ is the radius of the $(n-1)$-dimensional ball $\Bc^{n-1}(\mathbf{0}_n,\gamma)$. 
Note that the fact that $\gamma \in (0,1)$ follows from the structure of the original ellipsoid, i.e., $\left( K_\Nm^{-1}+I_n \right)^{-\frac{1}{2}}\mathcal{B}^{n}\left(\mathbf{0}_{n},1\right)$ since, by taking the eigendecomposition, we can write
\begin{align*}
\left( K_\Nm^{-1}+I_n \right)^{-\frac{1}{2}} = V (\Lambda^{-1} +I_n)^{-\frac{1}{2}} V^T,
\end{align*}
which implies $\gamma < 1$ since all elements of $(\Lambda^{-1} +I_n)^{-\frac{1}{2}}$ are strictly smaller than one.
We finally note that
\begin{align}
\Ec^n \cap \Wc 
& \overset{\rm (a)}{=} {\rm Proj}_{\Wc}(\Ec^n) \overset{\rm (b)}{=} {\rm Proj}_{\Wc}\left(\left( K_\Nm^{-1}+I_n \right)^{-\frac{1}{2}}\mathcal{B}^{n}\left(\mathbf{0}_{n},1\right) \right),
\end{align} where the labeled equalities follow from: ${\rm (a)}$ the fact that $\Ec^n$ is a convex set and is symmetric with respect to $\Wc$; and ${\rm (b)}$ the projection property of Steiner symmetrization in Proposition~\ref{prop:STprop}.
Thus,~\eqref{eq:intersectBall} becomes
\begin{align*}
{\rm Proj}_{\Wc}\left(\left( K_\Nm^{-1}+I_n \right)^{-\frac{1}{2}}\mathcal{B}^{n}\left(\mathbf{0}_{n},1\right) \right) = \Bc^{n-1}(\mathbf{0}_n,\gamma),
\end{align*}
where $\gamma \in (0,1)$.
This concludes the proof of Lemma~\thmref{BoundaryAt0}.
\end{IEEEproof}

\subsection{Proof of the Implication  $3) \Leftrightarrow 4)$} 
We here prove that  $3) \Leftrightarrow 4)$, namely we prove the following lemma.

\begin{lemma}
\label{lemma:KN}
Let $\Qc = \left \{ Q\in \mathcal{SO}(n):  \mathbf{q}_{n} = \frac{1}{\sqrt{n}}\mathbf{1}_n \right \}$, where $\mathcal{SO}(n)$ is the set of $n\times n$ real-valued orthonormal matrices, and $\mathbf{q}_{n}$ is the $n$-th column of $Q$.
Then, a $K_\Nm$ is a solution for Lemma~\thmref{BoundaryAt0}
if and only if there exists constants $(\gamma, a ,v)$ such that $\gamma\in(0,1)$, $a \in (0,1)$, $v\in\mathbb{R}$ satisfying $v^2 < \min \{a \gamma, (1-a)(1-\gamma)\}$~and
\begin{align*}
\left(K_\Nm^{-1}+I_n\right)^{-1} = Q \begin{bmatrix}
\gamma I_{n-2}  & 0_{n-2\times 2}
\\
0_{2\times n-2} & S
\end{bmatrix} Q^T,
\end{align*}
where $Q\in\Qc$ and $S=\left[ \begin{smallmatrix} \gamma & v \\ v & a \end{smallmatrix} \right]$.
\end{lemma}


\begin{IEEEproof}
%
We start by noting that any $n$-dimensional ellipsoid can be represented in terms of  a symmetric matrix.
In particular, an $n$-dimensional ellipsoid defined as $K^{\frac{1}{2}}\mathcal{B}^{n}\left(\mathbf{0}_{n},1\right)$ with $K$ being a positive definite matrix, can be equivalently represented as
\begin{align*}
K^{\frac{1}{2}}\mathcal{B}^{n}\left(\mathbf{0}_{n},1\right)&
=  \left \{\yv \in \mathbb{R}^n: \yv^T K^{-1} \yv \leq 1 \right \},
%
\end{align*} 
and hence
\begin{align*}
\left( K_\Nm^{-1}+I_n \right)^{-\frac{1}{2}}\mathcal{B}^{n}\left(\mathbf{0}_{n},1\right) = \left \{\xv \in\mathbb{R}^n :\xv^T \left( K_\Nm^{-1}+I_n \right) \xv \le 1 \right \}.
\end{align*}
Now, let $C$ be any $n  \times (n-1)$ matrix whose columns form an orthonormal basis of the hyperplane $\Wc=\{\xv \in \mathbb{R}^n:\mathbf{1}_n^T\xv=0 \}$, which is an $(n-1)$-dimensional vector subspace of $\mathbb{R}^n$. 
Then, from~\cite{karl1994reconstructing}, the relationship between the original ellipsoid $\left( K_\Nm^{-1}+I_n \right)^{-\frac{1}{2}}\mathcal{B}^{n}\left(\mathbf{0}_{n},1\right)$, which is specified by the matrix $\left( K_\Nm^{-1}+I_n \right)^{-1}$, and its projection on the hyperplane $\Wc$ namely ${\rm Proj}_{\Wc}(( K_\Nm^{-1}+I_n )^{-\frac{1}{2}}\mathcal{B}^{n}\left(\mathbf{0}_{n},1\right))$, which is specified by $B$ in the projection subspace, is given by the equation
\begin{align}
\label{eq:MatrProj}
B = C^T \left( K_\Nm^{-1}+I_n \right)^{-1}   C.
\end{align}
We want to find the necessary and sufficient conditions that ensure that the projection of the original ellipsoid $\left( K_\Nm^{-1}+I_n \right)^{-\frac{1}{2}}\mathcal{B}^{n}\left(\mathbf{0}_{n},1\right)$ on the hyperplane $\Wc$ is an $n-1$ dimensional ball, i.e., in~\eqref{eq:MatrProj} we need $B = \gamma I_{n-1}$, where $\gamma$
is the radius of the $n-1$ dimensional ball. We delegate the derivation of such necessary and sufficient conditions to Appendix~\ref{app:Matrix}.
\end{IEEEproof}

\subsection{Proof of the Implication  $4) \Rightarrow 5)$} 
We here prove that  $4) \Rightarrow 5)$, i.e., a $K_\Nm$ that satisfies Lemma~\ref{lemma:KN}
implies that $\Rc_{\pi,K_\Nm} = \left(K_\Nm + I_n \right) \Hc_\pi$, for all $\pi \in \mathcal{P}$.
Towards this end, we leverage the following auxiliary lemma, the proof of which is in Appendix~\ref{app:AuxRes}.
\begin{lemma}\label{lem:auxiliarity3)to4)} Let $\tilde{\Ym}_0  \sim \Nc \left( \mathbf{0}_n ,\tilde{K} \right)$ with  $\tilde{K}=\left( K_\Nm^{-1} + I_n \right)^{-1}$ that satisfies the condition in Lemma~\ref{lemma:KN}.  Then, there exists some $\beta \in (0,1)$ such that
\begin{align}
\Pr \left (\tilde{\Ym}_0 \in\Hc_\pi \right ) = \beta,~\forall\pi\in\Pc.   \label{eq:UniformityDueToK}
\end{align}
Moreover, if  $\tilde{\yv}\in\Hc_\tau$, then  
\begin{align}
\Pr \left (\tilde{\Ym}_0 + \tilde{\yv} \in \Hc_\tau \right ) = \max_{\pi\in\Pc} \left\{\Pr(\tilde{\Ym}_0 +  \tilde{\yv} \in  \Hc_\pi)\right\}. \label{eq:PrisMax}
\end{align} 
\end{lemma} 

We now leverage Lemma~\ref{lem:auxiliarity3)to4)} to prove the implication $4) \Rightarrow 5)$, and hence to conclude the proof of Theorem~\ref{thm:MainResult}. In particular, we have the following lemma.

\begin{lemma}\label{0Linear}
Suppose that $K_\Nm$ satisfies the conditions in Lemma~\ref{lemma:KN}. Then, 
\begin{align}
\Rc_{\pi,K_\Nm} = \left(K_\Nm + I_n \right) \Hc_\pi.
\end{align}
\end{lemma}

\begin{IEEEproof}
 Let $\tilde{\Ym}=  \tilde{\Ym}_0+\tilde{\yv}$ where $\tilde{\Ym}_0  \sim \Nc \left( \mathbf{0}_n ,\tilde{K} \right)$ with  $\tilde{K}=\left( K_\Nm^{-1} + I_n \right)^{-1}$, and  $\tilde{\yv}=\left( I_n + K_\Nm \right)^{-1}\yv$.   
Next, note that 
\begin{align}\label{eq:densityProb}
 f_\Ym(\yv,\Hc_\pi)
& = \int_{\xv\in\Hc_\pi} f_\Nm(\yv-\xv) f_\Xm(\xv)  \ {\rm d} \xv \nonumber \\
& = \int_{\xv\in\Hc_\pi} \frac{{\rm{e}}^{-\frac{1}{2}(\yv-\xv)^T K_\Nm^{-1} (\yv-\xv)}}{\sqrt{(2\pi)^n  \text{det}(K_\Nm)}}  \cdot \frac{{\rm{e}}^{-\frac{1}{2}\xv^T  \xv}}{\sqrt{(2\pi)^n }} \ {\rm d} \xv \nonumber \\
& = \int_{\xv\in\Hc_\pi} \frac{{\rm{e}}^{-\frac{1}{2}\left(\yv^T K_\Nm^{-1} \yv -2\yv^T K_\Nm^{-1} \xv + \xv^T (K_\Nm^{-1}+I_n) \xv \right) }}{(2\pi)^n \sqrt{\text{det}(K_\Nm) }} \ {\rm d} \xv \nonumber \\
& \overset{\rm (a)}{=} C_\yv \cdot \int_{\xv\in\Hc_\pi} \frac{ {\rm{e}}^{-\frac{1}{2}  (\tilde{\yv}-\xv)^T (K_\Nm^{-1} + I_n) (\tilde{\yv}-\xv) } 
}{\sqrt{(2\pi)^n \text{det}\left((K_\Nm^{-1} + I_n)^{-1}\right)}} \ {\rm d} \xv \nonumber \\
& \overset{\rm (b)}{=} C_\yv \cdot \Pr(\tilde{\Ym}\in\Hc_\pi) \nonumber \\
& = C_\yv \cdot \Pr(\tilde{\Ym}_0 + \tilde{\yv} \in\Hc_\pi),
\end{align} 
where the labeled equalities follow from: ${\rm (a)}$ defining
\begin{align*}
C_\yv & =\frac{\sqrt{\text{det}\left((K_\Nm^{-1} + I_n)^{-1}\right)} }{\sqrt{ (2\pi)^n \  \text{det}(K_\Nm) } } {\rm{e}}^{{-\frac{1}{2} \yv^TK_\Nm^{-1}\yv +\frac{1}{2} \tilde{\yv}^T(K_\Nm^{-1} + I_n)\tilde{\yv} }};
\end{align*}
and ${\rm (b)}$ noting that the integrand is equal to the multivariate Gaussian density $f_{\tilde{\Ym}}(\cdot)$.
%
%

Now if $\tilde{\yv}\in\Hc_\tau$ or equivalently if $\yv\in(K_\Nm+I_n)\Hc_\tau$, in view of \eqref{eq:densityProb} and using Lemma~\ref{lem:auxiliarity3)to4)}, we have that 
\begin{align}
f_\Ym(\yv,\Hc_\tau) 
& = C_\yv \cdot \Pr(\tilde{\Ym}_0 + \tilde{\yv} \in \Hc_\tau) \nonumber \\
& = C_\yv \cdot \max_{\pi\in\Pc} \left\{\Pr(\tilde{\Ym}_0 + \tilde{\yv} \in \Hc_\pi) \right\}\nonumber \\
& =  \max_{\pi\in\Pc} \left\{ C_\yv \cdot \Pr(\tilde{\Ym}_0 + \tilde{\yv} \in \Hc_\pi) \right\}\nonumber \\
& = \max_{\pi\in\Pc}\left\{ f_\Ym(\yv,\Hc_\pi) \right\}. 
\end{align} 
This indicates that  $\Hc_\tau$ is an optimal decision  for  all  $\yv\in(K_\Nm+I_n)\Hc_\tau$.  Consequently, when $K_\Nm$ satisfies the conditions in Lemma~\ref{lemma:KN}, we have that the optimal decision regions are given by
\begin{align}\label{eq:optlinear}
\Rc_{\pi,K_\Nm} = (K_\Nm+I_n)\Hc_\pi,~\forall\pi\in\Pc.
\end{align}
This concludes the proof of Lemma~\ref{0Linear}, and also of Theorem~\ref{thm:MainResult}.
\end{IEEEproof}

\section{Conclusion} 
\label{sec:Conclusion}
In this paper, we have considered a hypothesis testing framework to study a problem of data permutation recovery from an observation corrupted by correlated Gaussian noise. 
We have shown that the optimal decision regions may or may not be a linear transformation of the corresponding hypothesis regions depending on the noise covariance matrix.
We have focused on the linear regime, which is appealing from a computational perspective as within it the optimal decoding is of polynomial complexity in the data size.
We have characterized the optimal decision regions in the linear regime and showed that they are identical to the hypothesis of the observation multiplied by a permutation-independent linear function of the covariance matrix.
We have discussed several practical implications of this result. For instance, we have shown that when the data size is equal to two, the linear regime is the only regime, and when the data size is larger than two if the noise is memoryless then it must be isotropic to induce the linear regime.
By leveraging the structure of the optimal decision regions, we have also derived the probability of error in terms of a volume of a region that consists of the intersection of a cone with a linear transformation of the unit radius ball. 


\section*{Acknowledgments}
The authors are grateful to the anonymous reviewers for the especially careful reviews, which are reflected in the final version.

\appendices

\section{Proof of Proposition~\ref{lem:invPoint}}
\label{app:SymmDecReg}
We start by noting that any $\pi_1 \in\Pc$ has its own unique $\pi_2 \in\Pc$ such that $\Hc_{\pi_1}=-\Hc_{\pi_2}$ 
Then, for any observation $\yv$, we have that
\begin{align}\label{eq:InvPoint}
f_\Ym(\yv,\Hc_{\pi_1})
& = \int_{\xv\in\Hc_{\pi_1}} f_\Nm(\yv-\xv)f_\Xm(\xv) \ {\rm d} \xv \nonumber \\
& \overset{\rm (a)}{=} \int_{\zv\in-\Hc_{\pi_1}} f_\Nm(\yv+\zv)f_\Xm(\zv) \ {\rm d} \zv \nonumber \\
& \overset{\rm (b)}{=} \int_{\zv\in\Hc_{\pi_2}} f_\Nm(-\yv-\zv)f_\Xm(\zv) \ {\rm d} \zv \nonumber \\
& = f_\Ym(-\yv,\Hc_{\pi_2}),
\end{align} where the labeled equalities follow from: $\rm{(a)}$ change of variable $\zv=-\xv$; and $\rm{(b)}$ the fact that $\Hc_{\pi_1}=-\Hc_{\pi_2}$ and $f_\Nm(\nv)={f_\Nm(-\nv)}$. 

From the relation in~\eqref{eq:InvPoint}, it therefore follows that we can map $f_\Ym(\yv,\Hc_{\pi_1})$ to $f_\Ym(-\yv,\Hc_{\pi_2})$ for all $(\pi_1,\pi_2)$ index pairs where $\pi_1 \in \mathcal{P}$ and $\pi_2 \in \mathcal{P}$ such that $\Hc_{\pi_1}=-\Hc_{\pi_2}$.
Assume now that $\yv\in\Rc_{\pi_1,K_\Nm}$, which from~\eqref{region} implies that $f_\Ym(\yv,\Hc_{\pi_1})$ is the maximum among all $f_\Ym(\yv,\Hc_{\tau}),~\tau \in\Pc$. 
From~\eqref{eq:InvPoint} we then have that, among all $f_\Ym(-\yv,\Hc_{\tau}),~\tau \in\Pc$, the maximum joint density for $-\yv$ is $f_\Ym(-\yv,\Hc_{\pi_2})$ where $\pi_2$ is such that $\Hc_{\pi_2}=-\Hc_{\pi_1}$.
This, from~\eqref{region}, implies that
\begin{align}
-\yv\in\Rc_{\pi_2,K_\Nm}.
\end{align}
This concludes the proof of Proposition~\ref{lem:invPoint}.


\section{Proof of Proposition~\ref{prop:Spectrum}}
\label{app:EigenvaluesEigenvectors}
Let $\lambda_i,~i\in[1:n]$ be the eigenvalues of $K_\Nm$ and $\tilde{\lambda}_i,~i\in[1:n]$ be the eigenvalues of $(K_\Nm^{-1}+I_n)^{-1}$ in~\eqref{eq:ConditionOnKN}. Then, for all $i\in[1:n]$, the relationship between $\lambda_i$ and $\tilde{\lambda}_i$ is such that
\begin{align}
\label{eq:eigRel}
	\lambda_i = \frac{\tilde{\lambda}_i}{1-\tilde{\lambda}_i}.
\end{align}
For the case when $K_\Nm$ has $n$ equal eigenvalues (i.e., either $K_\Nm$ is a diagonal matrix with equal elements on the diagonal, or we take $v=0$ and $\gamma=a$ in~\eqref{eq:ConditionOnKN}), it is not difficult to verify that the eigenvalues and eigenvectors are of the form in~\eqref{eq:EigenFiniso}. 

We hence focus on the case when $K_\Nm$ has at least two distinct eigenvalues (i.e., the cases when either $v=0,\gamma\neq a$ or when $v\neq0$ in~\eqref{eq:ConditionOnKN}).
Since $(K_\Nm^{-1}+I_n)^{-1}$ in~\eqref{eq:ConditionOnKN} consists of an orthonormal matrix $Q \in \mathcal{Q}$ and a block diagonal matrix, its eigenvalues can be found as the solution of the following set of equations,
\begin{subequations}
\begin{align}
	\tilde{\lambda}_i &= \gamma,~i \in [1: n-2], \\ 
	a &= \tilde{\lambda}_{n-1} + \tilde{\lambda}_{n} - \gamma \label{eq:eigenTrace}, \\
	v^2 &= (\tilde{\lambda}_{n-1}-\gamma) (\gamma - \tilde{\lambda}_n ),
\end{align}
\end{subequations}
where the second expression is due to the fact that $\gamma+a = \tilde{\lambda}_{n-1} + \tilde{\lambda}_{n}$ and the last expression follows by computing the determinant of $S$ in~\eqref{eq:ConditionOnKN} with~\eqref{eq:eigenTrace}. By solving the above set of linear equations and by using~\eqref{eq:eigRel} we obtain the eigenvalues in~\eqref{eq:EigenvaluesFin} -- see also Appendix~\ref{app:Eigenvalues}.


We now use the eigenvalues in~\eqref{eq:EigenvaluesFin} to find the eigenvectors $\nuv_i$ of $K_\Nm$. We start by observing that $\nuv_i$'s are equal to the eigenvectors of $(K_\Nm^{-1}+I_n)^{-1}$.
Since the block matrix in~\eqref{eq:ConditionOnKN} has one isotropic matrix, we can easily find the first $n-2$ eigenvectors of $(K_\Nm^{-1}+I_n)^{-1}$ (i.e., those associated to the eigenvalue $\gamma$) as,
\begin{align}
	\nuv_i = \mathbf{q}_{i},~i \in [1: n-2],
\end{align} where $\mathbf{q}_{i}$ is $i$-th column of $Q\in\Qc$. 
For $\nuv_i,~i \in [n-1:n]$, by using the eigendecomposition of $S$ and the fact that $(K_\Nm^{-1}+I_n)^{-1} = QV{\tilde{\Lambda}}V^TQ^T$ with ${\tilde{\Lambda}}$ being a diagonal matrix and $V$ being an orthonormal matrix, we obtain the following two equations,
\begin{align}
	\nuv_i & = (a - \tilde{\lambda}_i) \mathbf{q}_{n-1} + v \mathbf{q}_{n}, i\in[n-1:n], \label{eq:eigVectwo1}\\
	\nuv_i & = v \mathbf{q}_{n-1} + (\gamma - \tilde{\lambda}_i) \mathbf{q}_{n}, i\in[n-1:n]. \label{eq:eigVectwo2}
\end{align} 
By combining~\eqref{eq:eigVectwo1} and \eqref{eq:eigVectwo2}, and by using~\eqref{eq:eigRel} we obtain the eigenvectors in~\eqref{eq:EigenvectorsFin}. 
This concludes the proof of Proposition~\ref{prop:Spectrum}.
%

\section{Proof of Proposition~\ref{prop:ProbError}}
\label{app:ErrorProb}
Instead of working with the probability of error, it is more convenient to work with the probability of correctness of our hypothesis testing problem. 
Using the structure of the optimal decision regions found in Theorem~\ref{thm:MainResult},  the probability of correctness can be written as 
\begin{align}
P_c 
& = \sum_{\pi  \in \mathcal{P}} \Pr \left(\left(\Xm,\Ym\right)^T\in \mathcal{H}_{\pi}\times \mathcal{R}_{\pi,K_{\mathbf{N}}} \right)\nonumber \\
& \stackrel{{\rm{(a)}}}{=} \sum_{\pi  \in \mathcal{P}} \Pr \left(\left(\Xm,\Ym\right)^T\in \mathcal{H}_{\pi}\times  (K_{\mathbf{N}}+I_n) \mathcal{H}_{\pi}\right ) \nonumber \\
& \stackrel{{\rm{(b)}}}{=} \sum_{\pi  \in \mathcal{P}}\Pr \left(\left(\Xm, \Xm + K_{\mathbf{N}}^{  \frac{1}{2} } {\mathbf{Z}} \right)^T\in \mathcal{H}_{\pi}\times  (K_{\mathbf{N}}+I_n) \mathcal{H}_{\pi}  \right) \notag \\
& \stackrel{{\rm{(c)}}}{=} \sum_{\pi  \in \mathcal{P}} \Pr \left(A(\Xm,{\mathbf{Z}})^T \in  \mathcal{H}_{\pi}\times  (K_{\mathbf{N}}+I_n) \mathcal{H}_{\pi}  \right)\nonumber \\
& \stackrel{{\rm{(d)}}}{=} \sum_{\pi  \in \mathcal{P}}  \Pr \left((\Xm,{\mathbf{Z}})^T \in A^{-1}  \mathcal{C}_{\mathcal{H}_{\pi}} \right) \notag \\
& \stackrel{{\rm{(e)}}}{=} n!   \Pr \left((\Xm,{\mathbf{Z}})^T \in A^{-1} \mathcal{C}_{\mathcal{H}_{\pi}}\right), \label{eq:ProbERR}
\end{align} 
where the labeled equalities follow from: 
$\rm{(a)}$ using the optimal decision regions  in Theorem~\ref{thm:MainResult};  
$\rm{(b)}$ letting $\mathbf{Z}$  be a standard normal random vector, i.e., $\mathbf{Z}\sim \mathcal{N}(\mathbf{0}_n,I_n)$;
 $\rm{(c)}$  defining $A=\begin{bmatrix} {I}_n& {0}_{n\times n}\\ {I}_n&  K_{\mathbf{N}}^{  \frac{1}{2} } \end{bmatrix} $;  
$\rm{(d)}$ letting $\mathcal{C}_{\mathcal{H}_{\pi}}=  \mathcal{H}_{\pi} \times  (K_{\mathbf{N}}+I_n) \mathcal{H}_{\pi}$; and $\rm{(e)}$ using the symmetry of  $(\Xm,\mathbf{Z})$.
 
 We observe that the shape of the region $\mathcal{H}_{\pi}$ is an $n$-dimensional cone (see Fig.~\ref{fig:RegN3} for a graphical representation when $n=3$). 
Thus, $\mathcal{C}_{\mathcal{H}_\pi}$ is a $2n$-dimensional cone and so is $ A^{-1} \mathcal{C}_{\mathcal{H}_{\pi}}$.      
It therefore follows that we have to determine the probability of  $(\Xm,{\mathbf{Z}})^T$ to fall within a cone.  Using the symmetry of the Gaussian distribution, the probability of a pair $(\Xm,{\mathbf{Z}})^T$  to fall within a cone is simply determined by the angular measure of the cone.   
Now, the angular measure of the cone  $A^{-1} \mathcal{C}_{\mathcal{H}_{\pi}}$ is given by 
\begin{align}
 \Pr \left((\Xm,{\mathbf{Z}})^T \in A^{-1} \mathcal{C}_{\mathcal{H}_{\pi}}\right) 
 &= \frac{ \mathrm{Vol}^{2n}\left(A^{-1}\mathcal{C}_{\mathcal{H}_\pi}\cap \mathcal{B}^{2n}\left(\mathbf{0}_{2n},1\right)\right)}{\mathrm{Vol}^{2n}\left(\mathcal{B}^{2n}\left(\mathbf{0}_{2n},1\right)\right)} \notag\\
  &= \frac{  | \mathrm{det}( A^{-1} ) |  \mathrm{Vol}^{2n}\left(\mathcal{C}_{\mathcal{H}_\pi}\cap A \mathcal{B}^{2n}\left(\mathbf{0}_{2n},1\right)\right)}{\mathrm{Vol}^{2n}\left(\mathcal{B}^{2n}\left(\mathbf{0}_{2n},1\right)\right)}, \label{eq:AngularMeasure}
\end{align}
where in the last equality we have used the fact that $ \mathrm{Vol}^k\left(A \mathcal{S} \right)= |\mathrm{det}(A ) | \mathrm{Vol}^k\left(\mathcal{S} \right)$ for any invertible matrix $A$ and any set $\mathcal{S}$. 
By combining~\eqref{eq:ProbERR} and~\eqref{eq:AngularMeasure} we arrive at 
\begin{align}
P_c = n! \frac{ |  \mathrm{det}( A^{-1} )  | \mathrm{Vol}^{2n}\left(\mathcal{C}_{\mathcal{H}_\pi}\cap A \mathcal{B}^{2n}\left(\mathbf{0}_{2n},1\right)\right)}{\mathrm{Vol}^{2n}\left(\mathcal{B}^{2n}\left(\mathbf{0}_{2n},1\right)\right)}. 
\end{align} 
The proof of Proposition~\ref{prop:ProbError} is concluded by noting that $A$ is a block matrix and hence  $ | \mathrm{det}( A ) | = \mathrm{det} \left ( K_{\mathbf{N}}^{  \frac{1}{2} } \right )$, and by using the fact that $P_e = 1- P_c$. 

\section{Proof of Lemma~\ref{lem:GaussianVolumue}}
\label{app:PrNinHpi}
We start by observing that, since $K_{\mathbf{U}}$ is positive definite, we have that
\begin{align}\label{eq:pr}
\Pr(\mathbf{U} \in\Hc_\pi)
& = \Pr \left (K_{\mathbf{U}}^{\frac{1}{2}}\Zm\in\Hc_\pi \right ) \nonumber \\
& = \Pr \left (\Zm\in K_{\mathbf{U}}^{-\frac{1}{2}} \Hc_\pi \right ) \nonumber \\
& =  \frac{ \mathrm{Vol}^n\left( K_{\mathbf{U}}^{-\frac{1}{2}} \Hc_\pi \cap \mathcal{B}^{n}\left(\mathbf{0}_{n},1\right)\right)}{\mathrm{Vol}^n\left(\mathcal{B}^{n}\left(\mathbf{0}_{n},1\right)\right)},
\end{align}
where $\Zm\sim\Nc(\mathbf{0}_n,I_n)$, and where the last equality follows by representing the probability in terms of a ratio of two volumes.
We then obtain
\begin{align}\label{eq:VolProb}
\mathrm{Vol}^n\left( K_{\mathbf{U}}^{-\frac{1}{2}} \Hc_\pi \cap \mathcal{B}^{n}\left(\mathbf{0}_{n},1\right)\right)
& = \left |{\rm det} \left (K_{\mathbf{U}}^{-\frac{1}{2}} \right ) \right | \mathrm{Vol}^n\left( \Hc_\pi \cap K_{\mathbf{U}}^{\frac{1}{2}} \mathcal{B}^{n}\left(\mathbf{0}_{n},1\right)\right),
\end{align} where the equality follows from the fact that, for an $n \times n$ invertible matrix $A$ and a set $\mathcal{S} \subseteq \mathbb{R}^n$, we have that $\mathrm{Vol}^n(A \mathcal{S}) = |{\rm det}(A)|\mathrm{Vol}^n(\mathcal{S})$.
%
Finally, by substituting~\eqref{eq:VolProb} into~\eqref{eq:pr} we obtain
\begin{align}\label{eq:Vol0_2}
\Pr(\mathbf{U}\in\Hc_\pi) 
& =  \frac{\left|{\rm det} \left (K_{\mathbf{U}}^{-\frac{1}{2}} \right) \right | \mathrm{Vol}^n\left( \Hc_\pi \cap K_{\mathbf{U}}^{\frac{1}{2}} \mathcal{B}^{n}\left(\mathbf{0}_{n},1\right)\right)}{\mathrm{Vol}^n\left(\mathcal{B}^{n}\left(\mathbf{0}_{n},1\right)\right)}.
\end{align}
This concludes the proof of Lemma~\ref{lem:GaussianVolumue}.

\section{Proof of Lemma~\ref{lem:conditionY0}}
\label{app:VolumeConsta}
We start by observing that, from the definition of the optimal decision regions in~\eqref{region}, we have that $\mathbf{0}_n \in \bigcap_{\pi \in \mathcal{P}}\Rc_{\pi,K_\Nm}  $ if and only if  
\begin{align}
f_\Ym(\mathbf{0}_n,\Hc_\pi)=d,~\forall \pi\in\Pc, \label{eq:NSconditionPDF0}
\end{align}
for some constant $d>0$.  
Note that this implies that 
\begin{align}\label{eq:Prdprime}
\Pr(\Xm\in\Hc_\pi | \Ym=\mathbf{0}_n ) = d^\prime,~\forall\pi\in\Pc,
\end{align} where $d^\prime = {d}/{f_\Ym(\mathbf{0}_n)}$.
Furthermore, recall that $\Xm|\Ym=\yv$ is Gaussian (see Remark~\ref{remarkConditional})  and 
\begin{align}\label{eq:VolProbStep}
\Pr\left(  {\bf X}\in \mathcal{H}_\pi | {\bf Y}\!=\!{\bf y}\right) \!=\! \Pr \left(   \left (I_n+K_{\bf N}^{-1} \right)^{-1}  {\bf y}  + \left (I_n+K_{\bf N}^{-1} \right)^{-\frac{1}{2}} {\bf Z}    \in \mathcal{H}_\pi  \right),\forall\pi\in\Pc,\yv\in\mathbb{R}^n,
\end{align} 
where ${\bf Z}$ is a standard Gaussian random vector. 


Now, by evaluating~\eqref{eq:VolProbStep} and combining it with Lemma~\ref{lem:GaussianVolumue}, we have that 
\begin{align}
\Pr(\Xm\in\Hc_\pi | \Ym=\mathbf{0}_n )
& =   \frac{ \left |{\rm det}(\mathring{K}^{-\frac{1}{2}}) \right | \mathrm{Vol}^n\left(  \Hc_\pi \cap \mathring{K}^{\frac{1}{2}}\mathcal{B}^{n}\left(\mathbf{0}_{n},1\right)\right)}{ \mathrm{Vol}^n\left(\mathcal{B}^{n}\left(\mathbf{0}_{n},1\right)\right)}, \label{eq:VOlumeRepresenationofPDF}
\end{align} where $\mathring{K}=\left(K_\Nm^{-1}+I_n \right)^{-1}$.
Finally, the sufficient and necessary condition in~\eqref{eq:Prdprime} together with~\eqref{eq:VOlumeRepresenationofPDF}, imply that
\begin{align}
    \frac{ \left |{\rm det}(\mathring{K}^{-\frac{1}{2}}) \right | \mathrm{Vol}^n\left(  \Hc_\pi \cap \mathring{K}^{\frac{1}{2}}\mathcal{B}^{n}\left(\mathbf{0}_{n},1\right)\right)}{\mathrm{Vol}^n\left(\mathcal{B}^{n}\left(\mathbf{0}_{n},1\right)\right)} = d^\prime,~\forall \pi \in \Pc,
\end{align} 
which, after rescaling and substituting $\mathring{K}=\left(K_\Nm^{-1}+I_n \right)^{-1}$, reduces to \eqref{eq:Vol0}
where 
\begin{align*}
\eta= d^\prime \frac{\mathrm{Vol}^n\left(\mathcal{B}^{n}\left(\mathbf{0}_{n},1\right)\right)}{\left |{\rm det}(\mathring{K}^{-\frac{1}{2}}) \right |}.
\end{align*}
This concludes the proof of Lemma~\ref{lem:conditionY0}.

\section{Proof of Lemma~\ref{n-1ball}}
\label{app:EllipsEqualRadii}

%
%
%
%
%
%
%
We start by noting that the proof of Lemma~\ref{n-1ball} for the case $n=2$ is immediate, and hence we next focus on the case $n>2$.
In particular, our proof will leverage an auxiliary result presented in the next lemma, the proof of which can be found in Appendix~\ref{app:lem:n-1vol:Proof}. 
\begin{lemma}\label{lem:n-1vol}
 Let $\Ec^n$ be an $n$-dimensional ellipsoid centered at the origin with unitary axes $\{\nuv_1,\nuv_2, \ldots,\nuv_n\}$ and corresponding radii equal to $\{r_1,r_2,  \ldots ,r_n\}$.  Moreover,  for  $r\in \mathbb{R}$, define the following hyperplane and $n-1$ dimensional ellipsoid:
\begin{align}
\Wc(r)&=\{\xv \in \mathbb{R}^n:\nuv_n^T\xv=r\},\\
\Ec^{n-1}_{\Wc(r)}&=\Ec^n \cap  \Wc(r).
\end{align}  
If $\nuv_n = \frac{1}{\sqrt{n}} \mathbf{1}_n$, then for every $\pi \in \mathcal{P}$
\begin{align}
{\rm Vol}^n\left( \Hc_\pi \cap \Ec^n \right)=  {\rm Vol}^{n-1}\left(\Hc_\pi \cap   \Ec_{\Wc(0)}^{n-1} \right) c(r_n),
\end{align} 
where $c(r_n)$ is a constant that only depends on $r_n$. 
\end{lemma} 
By leveraging Lemma~\ref{lem:n-1vol}, for a  constant $\eta>0$, we have that
\begin{align}
{\rm Vol}^n\left( \Hc_\pi \cap \Ec^n \right) =  \eta,~\forall\pi\in\Pc, \label{eq:Worigional}
\end{align} if and only if
\begin{align}\label{eq:EW0}
{\rm Vol}^{n-1}\left( \Hc_\pi \cap \Ec^{n-1}_{\Wc(0)} \right) = \tilde{\eta},~\forall\pi\in\Pc,
\end{align} where $\Ec^{n-1}_{\Wc(0)}=\Ec^{n}\cap\Wc(0)$, and where $\tilde{\eta}$ is some other constant.
Therefore, if~\eqref{eq:Worigional} holds then so does~\eqref{eq:EW0} and vice versa. 
Consequently, to prove Lemma~\ref{n-1ball}, we need to show that~\eqref{eq:EW0} holds if and only if $\Ec^{n-1}_{\Wc(0)}$ is an $(n-1)$-dimensional ball.
Remember that $\Ec^{n-1}_{\Wc(0)}\subset\Wc(0)$ has unitary axes $\{\nuv_1,\nuv_2,\ldots,\nuv_{n-1}\}$ with corresponding radii equal to $\{r_1,r_2,\ldots,r_{n-1}\}$. 

First, suppose that $\Ec^{n-1}_{\Wc(0)}$ is an $(n-1)$-dimensional ball. Then, from the symmetry of $\Hc_\pi$'s,  it readily follows that~\eqref{eq:EW0} holds (and hence~\eqref{eq:Worigional} holds). Hence, the fact that $\Ec^{n-1}_{\Wc(0)}$ is an $(n-1)$-dimensional ball is a sufficient condition for~\eqref{eq:Worigional} to hold. We now show that it is also necessary. In particular, our proof
follows by using a contradiction argument where we assume that $\Ec^{n-1}_{\Wc(0)}$ is not an $(n-1)$-dimensional ball.  

Assume that $\Ec^{n-1}_{\Wc(0)}$ has at least one radius that is different from the others. 
Without loss of generality, let $r_1=\max_{i \in [1:n-1]}\{r_i\}$ and   $r_2=\min_{i \in [1:n-1]}\{r_i\}$.
Note that $r_1\nuv_1\in\Ec_{\Wc(0)}^{n-1}$ and $r_2\nuv_2\in\Ec_{\Wc(0)}^{n-1}$.
Assume that $r_1\nuv_1\in\Hc_\alpha$ and $r_2\nuv_2\in\Hc_\beta$, for some  $\alpha,\beta\in\Pc$. 
Note that $\alpha \neq \beta$, i.e., when $n>2$, there is no possibility for any of the $\Hc_\pi$'s to contain more than one axis of $\Ec_{\Wc(0)}^{n-1}$.
Next, observe that  $\Hc_\alpha\cap\Wc(0)$ and $\Hc_\beta\cap\Wc(0)$ have equal $(n-1)$-dimensional cone shapes (i.e., the angular measures of the two cones are the same) in the subspace $\Wc(0)$.   
We let $\Bc^{n-1}_{\Wc}(\mathbf{0}_n,{r})=\Bc^{n}(\mathbf{0}_n,r)\cap \Wc(0)$ be the $(n-1)$-dimensional ball of radius $r$.
Because of the assumption of $r_1\neq r_2$, there exists some value $\tilde{r}$, such that $r_1>\tilde{r}>r_2$ and
\begin{align}\label{eq:ineqVol}
{\rm Vol}^{n-1}\left(\Hc_\alpha \cap   \Ec_{\Wc(0)}^{n-1} \right) 
& \overset{\rm (a)}{=} {\rm Vol}^{n-1}\left(\Hc_\alpha \cap \Wc(0) \cap   \Ec_{\Wc(0)}^{n-1} \right) \nonumber \\
& \overset{\rm (b)}{>} {\rm Vol}^{n-1}\left(\Hc_\alpha \cap \Wc(0) \cap   \Bc^{n-1}_{\Wc}(\mathbf{0}_n,\tilde{r}) \right) \nonumber \\
& \overset{\rm (c)}{=} {\rm Vol}^{n-1}\left(\Hc_\beta  \cap \Wc(0) \cap   \Bc^{n-1}_{\Wc}(\mathbf{0}_n,\tilde{r}) \right) \nonumber \\
& \overset{\rm (d)}{>} {\rm Vol}^{n-1}\left(\Hc_\beta \cap \Wc(0)  \cap   \Ec_{\Wc(0)}^{n-1} \right) \nonumber \\
& = {\rm Vol}^{n-1}\left(\Hc_\beta  \cap   \Ec_{\Wc(0)}^{n-1} \right),
\end{align}
where the labeled (in)equalities follow from: 
$\rm{(a)}$ the fact that $\Ec^{n-1}_{\Wc(0)}\subset\Wc(0)$;
$\rm{(b)}$ the assumption that the cone $\Hc_\alpha \cap \Wc(0)$ contains the largest axis of the ellipsoid (i.e., $r_1\nuv_1\in\Hc_\alpha$) and the assumption  $\tilde{r} <r_1$; $\rm{(c)}$ using the fact that  $\Hc_\beta, \Hc_\alpha, \Bc^{n-1}_{\Wc}(\mathbf{0}_n,\tilde{r})$ and $\Wc(0)$ are permutation invariant; and $\rm{(d)}$ the assumption that the cone $\Hc_\beta \cap \Wc(0)$ contains the smallest axis of the ellipsoid (i.e., $r_2\nuv_2\in\Hc_\beta$) and the assumption  $\tilde{r} >r_2$. 

This shows that, if $r_1 \neq r_2$, then~\eqref{eq:EW0} (and hence~\eqref{eq:Worigional}) can not hold.
Therefore, for~\eqref{eq:EW0} (and hence~\eqref{eq:Worigional}) to hold, $ \Ec_{\Wc(0)}^{n-1} $ must be an $(n-1)$-dimensional ball, i.e., the radii $\{r_1,\ldots,r_{n-1}\}$ of $\Ec^n$ must be all equal to each other. This concludes the proof of Lemma~\ref{n-1ball}.

\section{Necessary and Sufficient Conditions for Lemma~\ref{lemma:KN}}
\label{app:Matrix}
We start by noting that, by substituting $B = \gamma I_{n-1}$ inside~\eqref{eq:MatrProj}, we obtain
\begin{align}
\label{eq:MatrProj2}
&I_{n-1} \gamma = C^T \left( K_\Nm^{-1}+I_n \right)^{-1}   C.
\end{align}
Moreover, we also note that
\begin{align}\label{eq:CCT}
C C^T &\stackrel{{\rm{(a)}}}{=} 
\begin{bmatrix}
\mathbf{c}_1 & \mathbf{c}_2 & \ldots & \mathbf{c}_{n-1} 
\end{bmatrix}
\begin{bmatrix}
\mathbf{c}_1^T
\\
\mathbf{c}_2^T
\\
\vdots
\\
\mathbf{c}_{n-1}^T
\end{bmatrix}
\nonumber\\&
\stackrel{{\rm{(b)}}}{=} 
\begin{bmatrix}
\mathbf{c}_1 & \mathbf{c}_2 & \ldots & \mathbf{c}_{n} 
\end{bmatrix}
\left( 
I_n
-
\begin{bmatrix}
0_{(n-1)\times (n-1)} & \mathbf{0}_{n-1}
\\
\mathbf{0}_{n-1}^T & 1
\end{bmatrix}
\right )
\begin{bmatrix}
\mathbf{c}_1^T
\\
\mathbf{c}_2^T
\\
\vdots
\\
\mathbf{c}_{n}^T
\end{bmatrix} 
\nonumber\\& \stackrel{{\rm{(c)}}}{=}  I_n - \frac{1}{n} \mathbf{1}_n \mathbf{1}_n^T = I_n - \frac{1}{n} 1_{n\times n},
\end{align}
where the labeled equalities follow from: 
$\rm{(a)}$ letting $\mathbf{c}_i, i \in [1:n-1]$ be the $i$-th column of $C$;
$\rm{(b)}$ letting $\mathbf{c}_n = \frac{1}{\sqrt{n}} \mathbf{1}_n$;
and $\rm{(c)}$ noting that $\mathbf{c}_n$ is a unit vector that belongs to $\mathcal{L}_\Hc$ in~\eqref{eq:BoundHypReg} and hence, it is perpendicular to $\Wc$ and to its orthonormal basis formed by the $n-1$ columns of $C$.

Now recall that the set $\Qc$ is the set of $n\times n$ real-valued orthonormal matrices with the $n$-th column equal to $\frac{1}{\sqrt{n}}\mathbf{1}_n$. 
Moreover, note that since the matrix $C$ in~\eqref{eq:MatrProj2} is {\em any} orthonormal matrix the columns of which form a basis of the hyperplane $\Wc$, then the matrix $Q \in \mathcal{Q}$ can be chosen so as to have $C$ to populate its first $n-1$ columns. In other words, we can always find a pair $(Q,C)$ with $Q \in \Qc$ such that 
\begin{align}\label{eq:QCblock}
	Q = \begin{bmatrix} C & \frac{1}{\sqrt{n}}\mathbf{1}_n \end{bmatrix}.
\end{align}
Without loss of generality, we assume the structure in~\eqref{eq:QCblock} for $Q$, and we let
\begin{align}
\label{eq:DecKn}
\left(K_\Nm^{-1}+I_n\right)^{-1} = Q A Q^T.
\end{align}
Note that the matrix $A$ in~\eqref{eq:DecKn} is symmetric. This follows from the fact that the left-hand side of~\eqref{eq:DecKn} is positive definite, and hence symmetric. This implies that $Q A Q^T = (Q A Q^T)^T$, which leads to $A=A^T$.
Then, we obtain
\begin{align*}
C^T \left(K_\Nm^{-1}+I_n\right)^{-1} C = C^T Q A Q^T C =
\begin{bmatrix}
I_{n-1} & \mathbf{0}_{n-1}
\end{bmatrix}
A
\begin{bmatrix}
I_{n-1}
\\
\mathbf{0}_{n-1}^T
\end{bmatrix},
\end{align*}
and hence from~\eqref{eq:MatrProj2}, we need
\begin{align*}
\gamma I_{n-1} = 
\begin{bmatrix}I_{n-1} & \mathbf{0}_{n-1}
\end{bmatrix}
A
\begin{bmatrix}
I_{n-1}
\\
\mathbf{0}_{n-1}^T
\end{bmatrix},
\end{align*} 
which implies that $A$ has to have the form as
\begin{align*}
A = 
\begin{bmatrix}
\gamma I_{n-1}  & \mathbf{v}
\\
\mathbf{v}^T & a
\end{bmatrix},
\end{align*}
for some constant $a$ and column vector $\mathbf{v}$ of dimension $n-1$.
By substituting this back into~\eqref{eq:DecKn}, we obtain
\begin{align}
\label{eq:CondMatr}
\left(K_\Nm^{-1}+I_n\right)^{-1} = Q \begin{bmatrix}
\gamma I_{n-1}  & \mathbf{v}
\\
\mathbf{v}^T & a
\end{bmatrix} Q^T,
\end{align}
where $Q\in\Qc$.
Moreover, since we can arbitrarily choose the first $n-1$ columns of $Q\in\Qc$,
the expression in~\eqref{eq:CondMatr} can be further simplified as
\begin{align} 
\label{eq:CondMatr2}
\left(K_\Nm^{-1}+I_n\right)^{-1} 
&
= Q 
\begin{bmatrix}
\gamma I_{n-1}  & \mathbf{0}_{n-1}
\\
\mathbf{0}^T_{n-1} & a
\end{bmatrix} Q^T
+ Q \begin{bmatrix}
0_{n-1\times n-1}  & \mathbf{v}
\\
\mathbf{v}^T & 0
\end{bmatrix} 
Q^T \nonumber \\
&\overset{\rm (a)}{=} \tilde{Q} 
\begin{bmatrix}
\gamma I_{n-1}  & \mathbf{0}_{n-1}
\\
\mathbf{0}^T_{n-1} & a
\end{bmatrix} \tilde{Q}^T
+ \tilde{Q}  \begin{bmatrix}
0_{n-2\times n-2}  & 0_{n-2\times 2}
\\
0_{2\times n-2} & D
\end{bmatrix} 
\tilde{Q} ^T \nonumber \\
& = \tilde{Q}  \begin{bmatrix}
\gamma I_{n-2}  & 0_{n-2\times 2}
\\
0_{2\times n-2} & S
\end{bmatrix} 
\tilde{Q} ^T,
\end{align}
where $S=\left[\begin{smallmatrix} \gamma & v \\ v & a \end{smallmatrix} \right]$ with $v \in \mathbb{R}$, and where the equality in ${\rm (a)}$ follows since, for $Q \in \mathcal{Q}$ we have that 
\begin{align}
Q \begin{bmatrix}
0_{n-1\times n-1}  & \mathbf{v}
\\
\mathbf{v}^T & 0
\end{bmatrix} 
Q^T
& =  
\begin{bmatrix}
\cv_1 & \cdots & \cv_{n-1} & \frac{1}{\sqrt{n}}\mathbf{1}_n
\end{bmatrix}
\begin{bmatrix}
0_{n-1\times n-1}  & \mathbf{v}
\\
\mathbf{v}^T & 0
\end{bmatrix}
\begin{bmatrix}
\cv_1^T \\ \vdots \\ \cv_{n-1}^T \\ \frac{1}{\sqrt{n}}\mathbf{1}_n^T
\end{bmatrix} \nonumber \\
& \overset{\rm (a1)}{=} \frac{1}{\sqrt{n}}\mathbf{1}_n \left(\sum_{i=1}^{n-1} v_i \cv_{i}^T\right) + \left(\sum_{i=1}^{n-1} v_i \cv_i \right)\frac{1}{\sqrt{n}}\mathbf{1}_n^T  \nonumber \\
& \overset{\rm (a2)}{=}  \frac{v}{\sqrt{n}}\mathbf{1}_n \tilde{\cv}_{n-1}^T + \frac{v}{\sqrt{n}}  \tilde{\cv}_{n-1} \mathbf{1}_n^T \nonumber \\
& \overset{\rm (a3)}{=} \begin{bmatrix}
\tilde{\cv}_1 & \cdots & \tilde{\cv}_{n-1} & \frac{1}{\sqrt{n}}\mathbf{1}_n
\end{bmatrix}
\begin{bmatrix}
0_{n-2\times n-2}  & 0_{n-2\times n}
\\
0_{2\times n-2} & D
\end{bmatrix}
\begin{bmatrix}
\tilde{\cv}_1^T \\ \vdots \\ \tilde{\cv}_{n-1}^T \\ \frac{1}{\sqrt{n}}\mathbf{1}_n^T
\end{bmatrix} \nonumber \\
& \overset{\rm (a4)}{=} \tilde{Q} 
\begin{bmatrix}
0_{n-2\times n-2}  & 0_{n-2\times 2}
\\
0_{2\times n-2} & D
\end{bmatrix} 
\tilde{Q}^T,
\end{align}
where the labeled equalities follow from: 
$\rm(a1)$ letting $v_i,i \in [1:n-1]$ be the $i$-th element of $\vv$;
$\rm{(a2)}$ noting that we can express $\sum_{i=1}^{n-1} v_i\cv_i = v  \tilde{\cv}_{n-1}$ where $v$ is a scalar and $\tilde{\cv}_{n-1}\in\Wc$ is a unit vector orthogonal to $\frac{1}{\sqrt{n}}\mathbf{1}_n$;
$\rm{(a3)}$ the fact that $\tilde{\cv}_1, \ldots ,\tilde{\cv}_{n-1}$ is an orthonormal basis of the hyperplane $\Wc$ and using matrix form representation;
and
$\rm (a4)$ the fact that $\begin{bmatrix}
\tilde{\cv}_1 & \cdots & \tilde{\cv}_{n-1} & \frac{1}{\sqrt{n}}\mathbf{1}_n
\end{bmatrix} \in\Qc$, and letting $D = \left[ \begin{smallmatrix} 0 & v \\ v & 0 \end{smallmatrix} \right]$.

Thus, from~\eqref{eq:CondMatr2} we have that
\begin{align}
\label{eq:SimplEqCov}
\left(K_\Nm^{-1}+I_n\right)^{-1}  = \tilde{Q} \underbrace{\begin{bmatrix}
\gamma I_{n-2}  & 0_{n-2\times 2}
\\
0_{2\times n-2} & S
\end{bmatrix} }_{B} \tilde{Q}^T.
\end{align} 
Since $\left(K_\Nm^{-1}+I_n\right)^{-1}$ is a positive definite matrix, we need to ensure that the Schur complement~\cite{schur} of the block $\gamma I_{n-2}$ of the matrix $B$, denoted as $B/ \gamma I_{n-2}$, is positive definite. Formally,
\begin{align}
\label{eq:cond1}
B/ \gamma I_{n-2} = S \text{ is positive definite} \implies a \gamma > v^2 .
\end{align}
We also need to find the conditions that ensure that $K_\Nm$ is positive definite. Towards this end,
we perform the eigendecomposition of the matrix $B$, i.e., $B = V \Lambda V^T$, and rewrite~\eqref{eq:SimplEqCov} as
\begin{align}
\label{eq:SVDonA}
\left(K_\Nm^{-1}+I_n\right)^{-1} = Q V \Lambda V^T Q^T,
\end{align}
where we highlight that the matrix $QV$ is orthonormal.
Thus,
\begin{align}
&K_\Nm^{-1}+I_n  = (Q V \Lambda V^T Q^T)^{-1} = Q V \Lambda^{-1} V^T Q^T \nonumber
\\
\implies & K_\Nm^{-1} = Q V \Lambda^{-1} V^T Q^T - I_n = QV (\Lambda^{-1} - I_n) V^T Q^T \nonumber
\\
\implies & K_\Nm = QV (\Lambda^{-1} - I_n)^{-1} V^T Q^T.
\label{eq:CondKN}
\end{align}
In order to ensure that $K_\Nm$ is positive definite, we compute its eigenvalues, which are given by the diagonal elements of the diagonal matrix $(\Lambda^{-1} - I_n)^{-1}$ and we find the conditions under which these are positive. Note that these correspond to the conditions for which $\Lambda$ (i.e., the diagonal matrix with the eigenvalues of $B$) has diagonal elements strictly smaller than one.
The eigenvalues of $B$ are computed in Appendix~\ref{app:Eigenvalues}, where we have shown that $B$ has $n-2$ eigenvalues equal to $\gamma$ and the remaining two eigenvalues equal to 
\begin{align*}
\lambda = \frac{a+\gamma \pm \sqrt{(a-\gamma)^2 + 4 v^2}}{2}.
\end{align*}
These eigenvalues have to be strictly smaller than one, i.e., we need 
\begin{subequations}
\label{eq:cond2}
\begin{align}
\gamma < 1,
\end{align}
and
\begin{align}
\frac{a+\gamma \pm \sqrt{(a-\gamma)^2 + 4 v^2}}{2} <1 &\implies   \sqrt{(a-\gamma)^2 + 4 v^2} < 2-a-\gamma \nonumber
\\& \implies v^2 < (1-a)(1-\gamma).
\end{align}
\end{subequations}
Note also that since $v^2\geq 0$, we need $a < 1$.
The expression in~\eqref{eq:SimplEqCov} together with the conditions in~\eqref{eq:cond1} and~\eqref{eq:cond2} conclude the proof of Lemma~\ref{lemma:KN}.

\section{Proof of Lemma~\ref{lem:auxiliarity3)to4)}}
\label{app:AuxRes}

From the result in Lemma~\ref{lem:GaussianVolumue}, we have that 
 \begin{align*}
\Pr \left (\tilde{\Ym}_0 \in\Hc_\pi \right ) = \frac{\left |{\rm det} \left (\tilde{K}^{-\frac{1}{2}} \right ) \right | \mathrm{Vol}^n\left( \Hc_\pi \cap \tilde{K}^{\frac{1}{2}} \mathcal{B}^{n}\left(\mathbf{0}_{n},1\right)\right)}{\mathrm{Vol}^n\left(\mathcal{B}^{n}\left(\mathbf{0}_{n},1\right)\right)},  \ \forall \pi \in \mathcal{P},
\end{align*}
which together with Lemma~\ref{lem:conditionY0} lead to the proof of~\eqref{eq:UniformityDueToK}.  
Now, note that  \eqref{eq:UniformityDueToK} implies that 
\begin{align}
\beta= \Pr \left (\tilde{\Ym}_0 \in\Hc_\pi \right )= \Pr \left ({\bf Z} \in \tilde{K}^{-\frac{1}{2}} \Hc_\pi \right )  =\Pr \left ({\bf Z} \in \mathcal{C}_\pi \right ), \forall \pi \in \mathcal{P},
\end{align}
where ${\bf Z} \sim \mathcal{N}(\mathbf{0}_n,I_n)$
and  $\mathcal{C}_\pi=\tilde{K}^{-\frac{1}{2}} \Hc_\pi, \,  \forall \pi \in \mathcal{P}$.  
This further implies that  $\mathcal{C}_\pi, \pi \in \mathcal{P}$ is a collection of   congruent cones (i.e., cones with the same angular measure) that symmetrically partition $\mathbb{R}^n$.    
Moreover,  for every pair $(\tau, \pi) \in \mathcal{P} \times \mathcal{P}$ there exists a permutation matrix  $P_{\tau, \pi}$ such that $P_{\tau, \pi} \mathcal{C}_{\tau}=\mathcal{C}_\pi$
and
\begin{align}
\left \| {\bf x}  -  P_{\pi, \tau}  {\bf y}  \right \|  \le  \left \|   {\bf x}- {\bf y} \right \|,  \,   {\bf x} \in \mathcal{C}_\tau,  {\bf y} \in \mathcal{C}_\pi.   \label{eq:ConeInequality}
\end{align} 
The above inequality follows because of the three following facts: (i) $P_{\tau, \pi}  \mathcal{C}_{\tau}=\mathcal{C}_\pi$  implies that $\mathcal{C}_\pi$ is a reflection of $\mathcal{C}_{\tau}$ along some hyperplane $\mathcal{T}$; (ii) the hyperplane $\mathcal{T}$ bisects the distance between  $P_{\pi, \tau}  {\bf y}$ and ${\bf y}$ into equal segments;  and (iii) ${\bf x}$ and $ P_{\pi, \tau} {\bf y}$ are on the same side of the hyperplane  and ${\bf y}$ is on the opposite side of the hyperplane.
%
Therefore, the distance between  ${\bf x}$ and  $P_{\pi, \tau}  {\bf y}$ is smaller than the distance between ${\bf x}$ and ${\bf y}$.  

\begin{figure*}[t]
        \centering
%
%
\definecolor{mycolor1}{rgb}{0.24220,0.15040,0.66030}%
\definecolor{mycolor2}{rgb}{0.26470,0.40300,0.99350}%
\definecolor{mycolor3}{rgb}{0.10850,0.66690,0.87340}%
\definecolor{mycolor4}{rgb}{0.28090,0.79640,0.52660}%
\definecolor{mycolor5}{rgb}{0.91840,0.73080,0.18900}%
\definecolor{mycolor6}{rgb}{0.97690,0.98390,0.08050}%
\definecolor{mycolor7}{rgb}{0.00000,0.44700,0.74100}%
\definecolor{mycolor8}{rgb}{0.85000,0.32500,0.09800}%
\definecolor{mycolor9}{rgb}{0.92900,0.69400,0.12500}%
\begin{tikzpicture}

\begin{axis}[%
width=8cm,
height=8cm,
at={(3.844in,0.875in)},
scale only axis,
xmin=-1.2,
xmax=1.2,
ymin=-1.2,
ymax=1.2,
axis x line*=bottom,
axis y line*=left,
legend style={legend cell align=left, align=left, draw=white!15!black}
]

\addplot[area legend, draw=black]
table[row sep=crcr] {%
x	y\\
0	0\\
6.12323399573677e-17	1\\
-0.0615609061339428	0.998103328737044\\
-0.122888290664714	0.992420509671936\\
-0.18374951781657	0.982973099683902\\
-0.243913720108377	0.969796936035009\\
-0.303152674113043	0.952942000427157\\
-0.361241666187153	0.932472229404356\\
-0.417960344886783	0.908465271819524\\
-0.47309355683601	0.881012194285785\\
-0.526432162877356	0.850217135729614\\
-0.577773831408251	0.816196912356222\\
-0.626923805894106	0.779080574525671\\
-0.673695643646557	0.739008917220659\\
-0.717911923064442	0.696133945962927\\
-0.759404916654707	0.650618300204242\\
-0.798017227280239	0.602634636379256\\
-0.833602385221119	0.552364972960506\\
-0.866025403784438	0.5\\
0	0\\
}--cycle;


\addplot[area legend, draw=black]
table[row sep=crcr] {%
x	y\\
-0	0\\
-0.866025403784438	0.5\\
-0.895163291355062	0.445738355776539\\
-0.920905517944953	0.38978587329268\\
-0.943154434471277	0.33235479947966\\
-0.961825643172819	0.273662990072083\\
-0.976848317759601	0.213933083206498\\
-0.988165472081259	0.153391654878686\\
-0.995734176295034	0.0922683594633023\\
-0.999525719713366	0.0307950585561708\\
-0.999525719713366	-0.0307950585561701\\
-0.995734176295035	-0.0922683594633016\\
-0.988165472081259	-0.153391654878685\\
-0.976848317759601	-0.213933083206497\\
-0.961825643172819	-0.273662990072082\\
-0.943154434471278	-0.332354799479659\\
-0.920905517944954	-0.389785873292679\\
-0.895163291355063	-0.445738355776538\\
-0.866025403784439	-0.499999999999999\\
-0	0\\
}--cycle;


\addplot[area legend, draw=black]
table[row sep=crcr] {%
x	y\\
-0	-0\\
-0.866025403784439	-0.499999999999999\\
-0.83360238522112	-0.552364972960505\\
-0.79801722728024	-0.602634636379256\\
-0.759404916654708	-0.650618300204241\\
-0.717911923064442	-0.696133945962926\\
-0.673695643646558	-0.739008917220659\\
-0.626923805894107	-0.77908057452567\\
-0.577773831408252	-0.816196912356221\\
-0.526432162877356	-0.850217135729614\\
-0.473093556836011	-0.881012194285784\\
-0.417960344886784	-0.908465271819523\\
-0.361241666187154	-0.932472229404355\\
-0.303152674113045	-0.952942000427156\\
-0.243913720108378	-0.969796936035009\\
-0.183749517816571	-0.982973099683902\\
-0.122888290664715	-0.992420509671936\\
-0.0615609061339437	-0.998103328737044\\
-1.07187543957223e-15	-1\\
-0	-0\\
}--cycle;


\addplot[area legend, draw=black]
table[row sep=crcr] {%
x	y\\
-0	-0\\
-1.07187543957223e-15	-1\\
0.0615609061339416	-0.998103328737044\\
0.122888290664713	-0.992420509671936\\
0.18374951781657	-0.982973099683902\\
0.243913720108376	-0.96979693603501\\
0.303152674113042	-0.952942000427157\\
0.361241666187152	-0.932472229404356\\
0.417960344886782	-0.908465271819524\\
0.473093556836009	-0.881012194285785\\
0.526432162877355	-0.850217135729615\\
0.57777383140825	-0.816196912356222\\
0.626923805894105	-0.779080574525671\\
0.673695643646556	-0.73900891722066\\
0.717911923064441	-0.696133945962927\\
0.759404916654707	-0.650618300204243\\
0.798017227280239	-0.602634636379257\\
0.833602385221119	-0.552364972960507\\
0.866025403784438	-0.500000000000001\\
-0	-0\\
}--cycle;


\addplot[area legend, draw=black]
table[row sep=crcr] {%
x	y\\
0	-0\\
0.866025403784438	-0.500000000000001\\
0.895163291355062	-0.44573835577654\\
0.920905517944953	-0.389785873292681\\
0.943154434471277	-0.332354799479661\\
0.961825643172819	-0.273662990072084\\
0.9768483177596	-0.213933083206499\\
0.988165472081259	-0.153391654878687\\
0.995734176295034	-0.0922683594633038\\
0.999525719713366	-0.0307950585561714\\
0.999525719713366	0.0307950585561691\\
0.995734176295035	0.0922683594633006\\
0.98816547208126	0.153391654878684\\
0.976848317759601	0.213933083206496\\
0.96182564317282	0.273662990072081\\
0.943154434471278	0.332354799479659\\
0.920905517944954	0.389785873292678\\
0.895163291355063	0.445738355776537\\
0.866025403784439	0.499999999999999\\
0	-0\\
}--cycle;


\addplot[area legend, draw=black]
table[row sep=crcr] {%
x	y\\
0	0\\
0.866025403784439	0.499999999999999\\
0.833602385221121	0.552364972960504\\
0.798017227280241	0.602634636379255\\
0.759404916654708	0.650618300204241\\
0.717911923064443	0.696133945962925\\
0.673695643646559	0.739008917220658\\
0.626923805894108	0.779080574525669\\
0.577773831408253	0.81619691235622\\
0.526432162877357	0.850217135729613\\
0.473093556836011	0.881012194285784\\
0.417960344886785	0.908465271819523\\
0.361241666187155	0.932472229404355\\
0.303152674113045	0.952942000427156\\
0.243913720108378	0.969796936035009\\
0.183749517816572	0.982973099683901\\
0.122888290664716	0.992420509671936\\
0.0615609061339447	0.998103328737044\\
2.08251853918709e-15	1\\
0	0\\
}--cycle;

%
%

\draw [black,dashed,thick] (rel axis cs:0.57, 0.6) ellipse [x radius=45, y radius=45];

   \coordinate[label = left:$\boldsymbol{\mu}$] (B) at (rel axis cs:0.57,0.6) ;
   \node at (B)[circle,fill,inner sep=1.5pt]{};

\node
at (rel axis cs:0.6, 0.83) {$\mathcal{C}_{\{1,2,3\}}$};
\node
at (rel axis cs:0.3,0.7) {$\mathcal{C}_{\{1,3,2\}}$};
\node
at (rel axis cs:0.18,0.5) {$\mathcal{C}_{\{2,3,1\}}$};
\node
at (rel axis cs:0.4,0.2) {$\mathcal{C}_{\{2,1,3\}}$};
\node
at (rel axis cs:0.62,0.2) {$\mathcal{C}_{\{3,1,2\}}$};
\node
at (rel axis cs:0.83,0.5) {$\mathcal{C}_{\{3,2,1\}}$};

\end{axis}

\end{tikzpicture}%
        \caption{A pictorial depiction of the inequality in~\eqref{eq:PermuationInequalities} for $n=3$ and $\tau=\{1,2,3\}$.}
        \label{fig:PieExampleLemma7}
    \end{figure*}
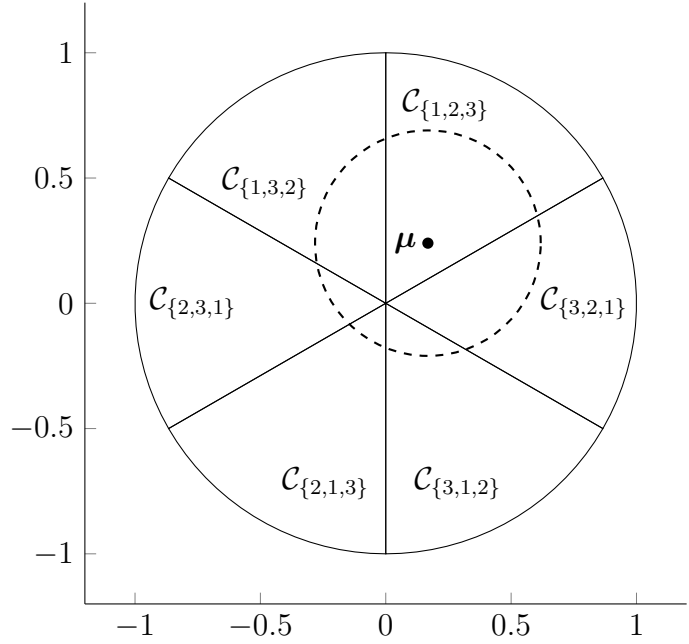

Next, with some abuse of notation, we let $f_{\bf Z}( \|  {\bf z}\|  )$ denote the PDF of ${\bf Z}$. This notation highlights the fact that the PDF of ${\bf Z}$  only depends on the norm.  We also define ${\bf \muv}=\tilde{K}^{-\frac{1}{2}}  \tilde{\yv} $ where ${ \bf \muv} \in \mathcal{C}_{\tau}$ since by assumption $\tilde{\yv} \in \Hc_\tau$. With this, we obtain
  \begin{align}
\Pr \left (\tilde{\Ym}_0 +  \tilde{\yv} \in \Hc_\pi \right )  & = \Pr \left ( \mathbf{Z}  +\left( K_\Nm^{-1} + I_n \right)^{1/2} \tilde{\yv} \in \left( K_\Nm^{-1} + I_n \right)^{1/2}  \Hc_\pi \right ) \nonumber \\
& \stackrel{{\rm{(a)}}}{=}  \Pr( {\bf Z}  + \boldsymbol{\mu} \in \mathcal{C}_{\pi}) \nonumber \\
&=  \int_{\mathcal{C}_{\pi}}  f_{\bf Z}( \|  {\bf z}- {\boldsymbol \mu} \|  )\  {\rm d} {\bf z} \nonumber \\
& \stackrel{{\rm{(b)}}}{\leq}  \int_{\mathcal{C}_{\pi}}  f_{\bf Z}( \|  P_{\pi,\tau} {\bf z} - {\boldsymbol \mu} ) \|  ) \ {\rm d} {\bf z} \nonumber \\
& \stackrel{{\rm{(c)}}}{=}  \int_{ P_{\pi,\tau} \mathcal{C}_{\pi} }  f_{\bf Z}( \|  {\bf z}- \boldsymbol{\mu} ) \|  ) \ {\rm d} {\bf z} \nonumber \\
&\stackrel{{\rm{(d)}}}{=} \int_{ \mathcal{C}_{\tau} }  f_{\bf Z}( \|  {\bf z}- \boldsymbol{\mu} ) \|  ) \ {\rm d} {\bf z} \nonumber \\
&=  \Pr \left ( {\bf Z}  + \boldsymbol{\mu} \in \mathcal{C}_{\tau} \right) \nonumber\\
&=\Pr \left (\tilde{\Ym}_0 +  \tilde{\yv} \in \Hc_\tau \right ),  \label{eq:PermuationInequalities}
\end{align}  
where the labeled (in)equalities follow from:
$\rm{(a)}$ letting $\boldsymbol{\mu}= \left( K_\Nm^{-1} + I_n \right)^{1/2} \tilde{\yv}$ and remembering that $ \mathcal{C}_{\pi} = \tilde{K}^{-\frac{1}{2}} \Hc_\pi = \left( K_\Nm^{-1} + I_n \right)^{1/2}  \Hc_\pi$ for all $\pi \in \mathcal{P}$; $\rm{(b)}$ applying the bound in~\eqref{eq:ConeInequality} and noting that $\boldsymbol{\mu} \in \mathcal{C}_{\tau}$; $\rm{(c)}$ using change of variable and the fact that $|{\rm det}(P_{\tau,\pi})|=1$;
and $\rm{(d)}$ the fact that $\mathcal{C}_{\tau} = P_{\pi, \tau} \mathcal{C}_{\pi}$.
The geometric interpretation of the inequality in $\rm{(b)}$ is shown in Fig.~\ref{fig:PieExampleLemma7}. 
In particular, in Fig.~\ref{fig:PieExampleLemma7} the view is taken with respect to the axis of symmetry.  The dashed ball centered at $\boldsymbol{\mu}$  is meant to  represent a level set of the PDF of ${\bf Z}+\boldsymbol{\mu}$. The intersection of the dashed ball and a cone $\mathcal{C}_\pi$ is the largest for the cone in which $\boldsymbol{\mu}$ lies, i.e., $\pi = \{1,2,3\}$. The proof of Lemma~\ref{lem:auxiliarity3)to4)} is concluded  by noting that~\eqref{eq:PermuationInequalities} holds with equality if $\tau=\pi$.
%

\section{Proof of Lemma~\ref{lem:n-1vol}} 
\label{app:lem:n-1vol:Proof}
Let $\Ec^n$ be an $n$-dimensional ellipsoid centered at the origin with unitary axes $\{\nuv_1,\nuv_2, \ldots,\nuv_n\}$ and corresponding radii equal to $\{r_1,r_2,  \ldots ,r_n\}$. 
Let one of the axes of $\Ec^n$ be equal to $\frac{1}{\sqrt{n}} \mathbf{1}_n$.
Specifically, without loss of generality, we set $\nuv_n = \frac{1}{\sqrt{n}} \mathbf{1}_n$, which has $r_n$ as corresponding radius.
Then, by introducing the hyperplane $\Wc(r)=\{\xv\in\mathbb{R}^n:\nuv_n^T\xv=r\}$, for any $\pi \in \mathcal{P}$, we can represent the volume of the intersection between $\Hc_\pi$ and $\Ec^n$ as
\begin{align}\label{eq:intvol}
{\rm Vol}^n\left( \Hc_\pi \cap \Ec^n \right)
& = \int_{-r_n}^{r_n} {\rm Vol}^{n-1}\left(\Hc_\pi \cap \Ec^n \cap  \Wc(r) \right) {\rm d}r \nonumber\\
& = \int_{-r_n}^{r_n} {\rm Vol}^{n-1}\left(\Hc_\pi \cap \Ec_{\Wc(r)}^{n-1} \right) {\rm d}r,
\end{align} 
where 
$\Ec_{\Wc(r)}^{n-1}=\Ec^n \cap  \Wc(r) $ is an $(n-1)$-dimensional ellipsoid in $\mathbb{R}^{n}$.

Note that since $\Ec^{n-1}_{\Wc(r)}$ has $\nuv_n$ as normal vector, which is one of the axes of $\Ec^n$, the ellipsoid $\Ec_{\Wc(r)}^{n-1}$ can be represented as
\begin{align}\label{eq:sameE}
\Ec_{\Wc(r)}^{n-1}
& = m(r) I_{n}\cdot \Ec_{\Wc(0)}^{n-1} + r\nuv_n,
\end{align} where $m(r):[-r_n,r_n]\rightarrow (0,1]$ is some magnitude function.
Then, we have
\begin{align}\label{eq:volind}
{\rm Vol}^n\left( \Hc_\pi \cap \Ec^n \right)
& \overset{\rm (a)}{=} \int_{-r_n}^{r_n} {\rm Vol}^{n-1}\left(\Hc_\pi \cap \left\{ m(r) I_n\cdot \Ec_{\Wc(0)}^{n-1} + r\nuv_n \right\}\right) {\rm d}r \nonumber \\
& \overset{\rm (b)}{=} \int_{-r_n}^{r_n} {\rm Vol}^{n-1}\left(\Hc_\pi \cap m(r) I_n\cdot \Ec_{\Wc(0)}^{n-1} \right) {\rm d}r \nonumber \\
& \overset{\rm (c)}{=}  \int_{-r_n}^{r_n} | {\rm det}\left(m(r) I_{n}\right) | {\rm Vol}^{n-1}\left(\Hc_\pi \cap   \Ec_{\Wc(0)}^{n-1} \right) {\rm d}r \nonumber \\
& ={\rm Vol}^{n-1}\left(\Hc_\pi \cap   \Ec_{\Wc(0)}^{n-1} \right)  \int_{-r_n}^{r_n} m(r)^{n}   {\rm d}r,
\end{align} where the labeled equalities follow from: $\rm{(a)}$ substituting~\eqref{eq:sameE} into~\eqref{eq:intvol}; 
$\rm{(b)}$ the fact that $\Hc_\pi,~\forall\pi \in \mathcal{P}$ is invariant to adding $a\nuv_n$,  where $a\in\mathbb{R}$ is any constant and remember that $\nuv_n = \frac{1}{\sqrt{n}} \mathbf{1}_n$ (i.e., $\Hc_\pi=\Hc_\pi + a\nuv_n$); and $\rm{(c)}$ the facts that, for any invertible matrix $A$ and any set $\mathcal{S}$, ${\rm Vol}^n\left( A\Sc\right)=| {\rm det}\left( A\right) |{\rm Vol}^n\left( \Sc\right)$ and $\Hc_\pi=k I_n\Hc_\pi$, where $k$ is any positive number.
We conclude the proof of Lemma~\ref{lem:n-1vol} by defining $c(r_n)=\int_{-r_n}^{r_n} m(r)^{n}   {\rm d}r$.

\section{Eigenvalues of $B$ in~\eqref{eq:SimplEqCov}}
\label{app:Eigenvalues}
We seek to compute the eigenvalues of the matrix $B$ defined as
\begin{align}
\label{eq:AMatr}
B = \begin{bmatrix}
\gamma I_{n-2}  & 0_{n-2\times 2}
\\
0_{2\times n-2} & S
\end{bmatrix},
\end{align}
where $S = \left[ \begin{smallmatrix} \gamma & v \\ v & a \end{smallmatrix}\right]$ is a $2\times 2$ symmetric matrix.
These can be found as the values of $\lambda$ that satisfy the equation
\begin{align*}
\text{det}(B- \lambda I_n) = 0 &\implies \text{det} \left ( \begin{bmatrix}
(\gamma-\lambda) I_{n-2}  & 0_{n-2\times 2}
\\
0_{2\times n-2} & S-\lambda I_{2}
\end{bmatrix}\right ) = 0
\\& \implies \text{det} \left ( (\gamma-\lambda) I_{n-2} \right ) \text{det} \left ( S -\lambda I_2 \right ) = 0
%
%
\\& \implies (\gamma-\lambda)^{n-2} ((a-\lambda)(\gamma-\lambda) - v^2) = 0.
\end{align*}
Hence the matrix $B$ in~\eqref{eq:AMatr} has $n-2$ eigenvalues equal to $\gamma$ and the remaining two eigenvalues can be found as the solution of
\begin{align}
 (a-\lambda)(\gamma-\lambda) -  v^2 = 0 \
 \implies \ & \lambda = \frac{a+\gamma \pm \sqrt{(a-\gamma)^2 + 4 v^2}}{2}.
\label{eq:EigNottrivial}
\end{align}

\bibliographystyle{IEEEtran}
\bibliography{HypoTestNetwork}

\end{document}